\documentclass{aa}
\usepackage{graphicx}
\usepackage{txfonts}
\usepackage{epsfig}
\usepackage{natbib}

\usepackage{pifont}

\newcommand{\cmark}{\ding{51}}%
\newcommand{\xmark}{\ding{55}}%

\begin{document} 

\title{Hydrodynamic simulations unravel the progenitor-supernova-remnant
connection in SN\,1987A}

\author{S.\ Orlando\inst{1}
   \and M.\ Ono\inst{2,3}
   \and S.\ Nagataki\inst{2,3}
   \and M.\ Miceli\inst{4,1}
   \and H.\ Umeda\inst{5}
   \and G.\ Ferrand\inst{2,3}
   \and F.\ Bocchino\inst{1} \and \\
   O.\ Petruk\inst{6,7}
   \and G.\ Peres\inst{4,1}
   \and K.\ Takahashi\inst{8}
   \and T.\ Yoshida\inst{5}
}

\offprints{S. Orlando}

\institute{INAF -- Osservatorio Astronomico di Palermo, Piazza del Parlamento 1, I-90134 Palermo, Italy\\ 
\email{salvatore.orlando@inaf.it}
\and Astrophysical Big Bang Laboratory, RIKEN Cluster for Pioneering Research, 2-1 Hirosawa, Wako, Saitama 351-0198, Japan
\and RIKEN Interdisciplinary Theoretical \& Mathematical Science Program (iTHEMS), 2-1 Hirosawa, Wako, Saitama 351-0198, Japan
\and Dip. di Fisica e Chimica, Universit\`a degli Studi di Palermo, Piazza del Parlamento 1, 90134 Palermo, Italy
\and Department of Astronomy, Graduate School of Science, University of Tokyo, Tokyo 113-0033, Japan
\and Institute for Applied Problems in Mechanics and Mathematics, Naukova Street 3-b, Lviv 79060, Ukraine
\and Astronomical Observatory, National University, Kyryla and Methodia St 8, Lviv 79008, Ukraine
\and Max-Planck-Institut f\"ur Gravitationsphysik (Albert-Einstein-Institute), Am M\"uhlenberg 1, D-14476 Potsdam-Golm, Germany
  }

\date{Received date / Accepted date}

\abstract
%Context
{Massive stars end their lives with catastrophic supernova (SN)
explosions. Key information on the explosion processes and on the
progenitor stars can be extracted from observations of supernova
remnants (SNRs), the outcome of SNe. Deciphering these observations
however is challenging because of the complex morphology of SNRs.}
%Aims
{We aim at linking the dynamical and radiative properties of the
remnant of SN\,1987A to the geometrical and physical characteristics
of the parent aspherical SN explosion and to the internal structure
of its progenitor star.}
%Methods
{We performed comprehensive three-dimensional hydrodynamic simulations
which describe the long-term evolution of SN 1987A from the onset
of the SN to the full-fledged remnant at the age of 50 years, accounting
for the pre-SN structure of the progenitor star. The simulations
include all physical processes relevant for the complex
phases of SN evolution and for the interaction of the SNR with
the highly inhomogeneous ambient environment around SN 1987A.
Furthermore the simulations follow the life cycle of elements from
the synthesis in the progenitor star, through the nuclear reaction
network of the SN, to the enrichment of the circumstellar medium
through mixing of chemically homogeneous layers of ejecta. From the
simulations, we synthesize observables to be compared with
observations.}
%Results
{By comparing the model results with observations, we constrained
the initial SN anisotropy causing Doppler shifts observed in emission
lines of heavy elements from ejecta, and leading to the remnant
evolution observed in the X-ray band in the last thirty years. In
particular, we found that the high mixing of ejecta unveiled by
high redshifts and broadenings of [Fe\,II] and $^{44}$Ti lines
require a highly asymmetric SN explosion channeling a significant
fraction of energy along an axis almost lying in the plane of the
central equatorial ring around SN\,1987A, roughly along the
line-of-sight but with an offset of $40^{\rm o}$, with the lobe
propagating away from the observer slightly more energetic than the
other. Furthermore, we found unambiguously that the observed
distribution of ejecta and the dynamical and radiative properties
of the SNR can be best reproduced if the structure of the progenitor
star was that of a blue supergiant resulted from the merging of two
massive stars.}
%Conclusions
{}


\titlerunning{Hydrodynamic simulations unravel the
progenitor-supernova-remnant connection in SN\,1987A}
\authorrunning{S. Orlando et~al.}

\maketitle

\section{Introduction}
\label{sec:intro}

The structure and morphology of supernova remnants (SNRs), the
outcome of supernova (SN) explosions, reflect the properties
of the parent SNe and the characteristics of their stellar progenitor
systems. Thus establishing firm connections between SNRs and their
parent SNe and progenitor stars is essential to obtain key
information on many aspects of the explosion processes associated
with SNe and of the final phases of stellar evolution. A unique
opportunity to study these connections is offered by the remnant
of SN\,1987A in the Large Magellanic Cloud, thanks to its youth
($\approx 30$~years) and proximity ($\approx 50$~kpc).

Observations of this remnant in different wavelength bands and at
different epochs have revealed Doppler shifts in emission lines of
heavy elements (e.g. [Fe\,II] and [Ni\,II]) up to velocities $>
3000$~km~s$^{-1}$ (e.g. \citealt{1990ApJ...360..257H, 1994ApJ...427..874C,
1995A&A...295..129U, 2016ApJ...833..147L}), suggesting an aspherical
explosion and the production of high-velocity clumps of metal-rich
ejecta in the SN. This scenario is also supported by the detection
of emission lines from decay of $^{44}$Ti around day $\approx 10000$
(\citealt{2015Sci...348..670B}) which are redshifted with Doppler
velocities of $\approx 700\pm 400$~km~s$^{-1}$. All these lines of
evidence point to large-scale asymmetry in the SN explosion, possibly
a signature of the engine powering the burst.

Identifying the physical and geometrical properties of initial SN
asymmetries is crucial to constrain explosion processes associated
with SNe. Seminal studies (e.g. \citealt{2015ApJ...803..101P,
2015A&A...577A..48W, 2015ApJ...810..168O, 2016ApJ...822...22O,
2017ApJ...842...13W, 2019ApJ...877..136F}) have shown that
hydrodynamic/magnetohydrodynamic (HD/MHD) models can be very effective
in studying the stellar progenitor-SN-SNR connection. Current models,
however, treat the problem in a piecemeal way mainly due to the
very different time and space scales involved, and the intrinsic
three-dimensional (3D) nature of the phenomenon. Additional difficulty
stems from the need to disentangle, in the remnant morphology, the
effects of the SN explosion and of the structure of the stellar
progenitor, from those of the interaction of the blast with the
inhomogeneous ambient environment. As a result, many aspects of
the stellar progenitor-SN-SNR connection remain uncertain.

Here we link the dynamical and radiative properties of the remnant
of SN\,1987A to the asymmetric SN explosion and to the structure
of the progenitor star, through a comprehensive 3D hydrodynamic
model which describes the evolution from the onset of the SN to the
development of its remnant, accounting for the pre-SN structure of
the progenitor star. To capture the large range in space scales
(spanning 13 orders of magnitude) during the evolution, we coupled
a 3D model of a core-collapse SN (\citealt{ono2019matter}, in the
following Paper I) with a 3D model of a SNR
(\citealt{2019A&A...622A..73O}).

The paper is organized as follows. In Sect.~\ref{sec:model} we
describe the model, the numerical setup, and the synthesis of X-ray
emission; in Sect.~\ref{results} we discuss the results; and in
Sect.~\ref{conclusion} we draw our conclusions.

\section{Problem description and numerical setup}
\label{sec:model}

We described the evolution of SN\,1987A from the onset of the SN
to the full-fledged remnant following a strategy analogous to that
outlined in previous works (\citealt{2015ApJ...810..168O,
2016ApJ...822...22O}). First we performed 3D core-collapse SN
simulations (see Sect.~\ref{SNmodel}) which describe the onset of
the SN, the propagation of the shock wave through the stellar
interior, and the breakout of the shock wave at the stellar surface.
Thus, the model accounts for the mixing and clumping of
the stellar envelope and mantle, i.e. the ejecta, at the time of
shock breakout. The SN simulations were initialized from pre-SN
stellar models available in the literature (see Sect.~\ref{stellar_prog})
and they ended soon after the shock breakout, covering $\approx 20$
hours of evolution. Then, the output of these simulations was used
as initial condition for the structure and chemical composition of
ejecta in 3D MHD simulations which describe the subsequent evolution:
the transition from the phase of SN to that of SNR and the interaction
of the remnant with the inhomogeneous nebula around SN\,1987A (see
Sect.~\ref{SNRmodel}). The SNR simulations cover 50 years, thus
describing the past evolution of SN\,1987A until its current age
and predicting its future evolution (until 2037).

\subsection{Pre-supernova models}
\label{stellar_prog}

\begin{figure*}[!t]
  \begin{center}
  \leavevmode
  \epsfig{file=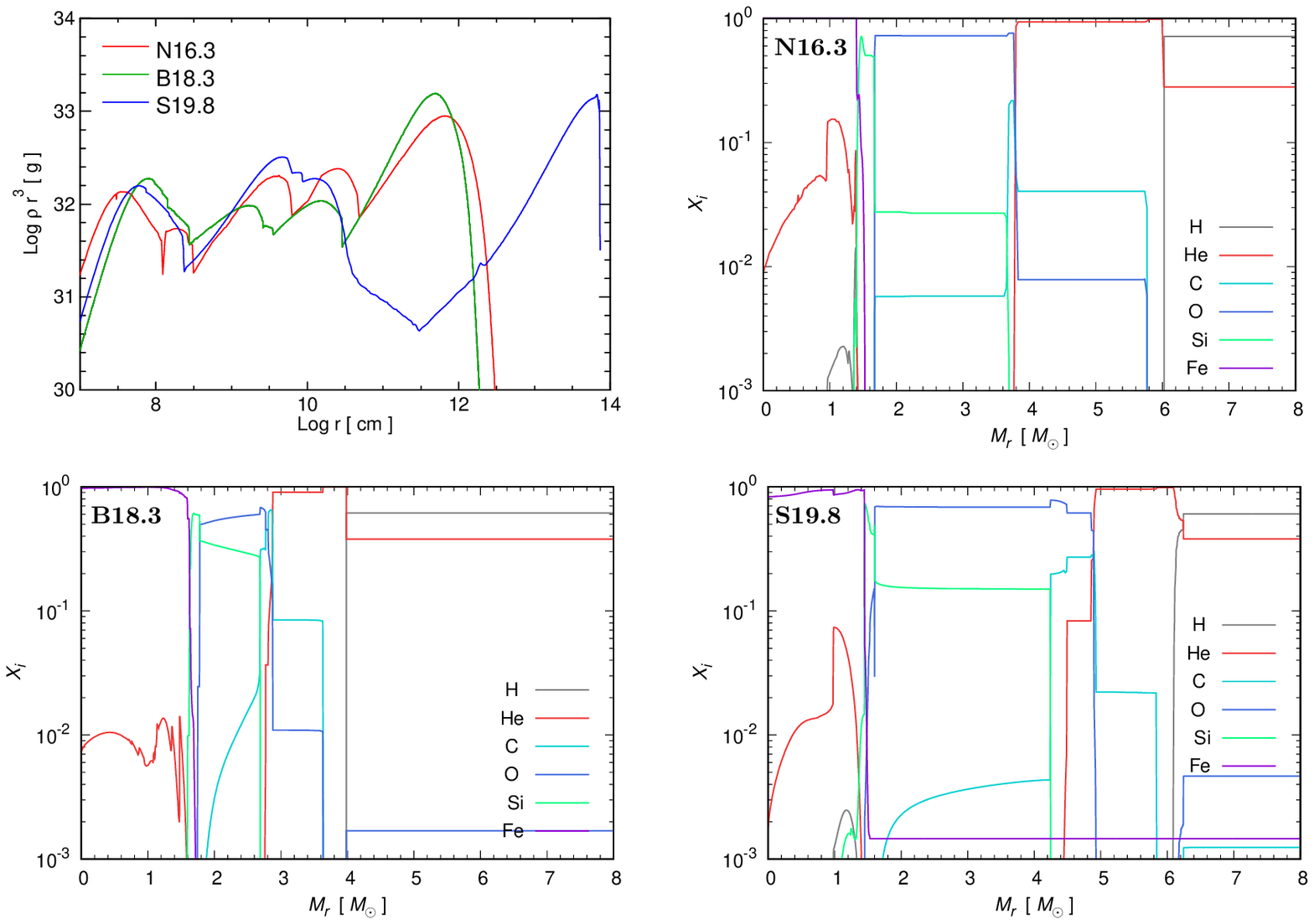, width=17.cm}
  \caption{Radial profiles of $\rho r^3$ of the pre-SN models
  considered in this paper (upper left panel) and mass fractions
  of selected elements (see legend in each panel) versus the
  interior mass $M_r$ in the pre-SN models N16.3 (upper right;
  \citealt{1988PhR...163...13N, 1990ApJ...360..242S}), B18.3 (lower
  left; \citealt{2018MNRAS.473L.101U}), and S19.8 (lower right;
  \citealt{2002RvMP...74.1015W, 2016ApJ...821...38S}).}
\label{star_prog}
\end{center}
\end{figure*}

The progenitor star of SN\,1987A, Sanduleak (Sk) $-69^{\rm o}\, 202$,
was a blue supergiant (BSG) with an initial mass of $\approx
20\,M_{\odot}$ (\citealt{1987Natur.327..597H}). In the present study,
we considered two progenitor star models of BSG (N16.3 and B18.3),
which have been proposed in the literature to describe the average
radial structure of Sk $-69^{\rm o}\, 202$ at collapse, and a model
of red supergiant (RSG; model S19.8) selected to explore the evolution
and structure of the remnant in the case of a progenitor star
which differs significantly from the structure of a BSG.
Figure~\ref{star_prog} shows the radial profiles of $\rho r^3$
(where $\rho$ is the mass density, and $r$ is the radial distance
from the center of the star) and the mass fractions of selected
elements versus the interior mass, $M_r$, of the pre-SN models
considered here. The radial profile of $\rho r^3$ provides useful
information on the SN shock dynamics because its gradient determines
where and when the SN blast wave decelerates (positive
gradient) or accelerates (negative gradient).

The first model, N16.3, denotes a $16.3 M_{\odot}$ BSG
that was early proposed to describe Sk $-69^{\rm o}\, 202$
(\citealt{1988PhR...163...13N, 1990ApJ...360..242S}). The model assumes a
main-sequence mass of $20 M_{\odot}$ which reduces to $16.3 M_{\odot}$
at the onset of the SN due to mass loss. For this reason, the model is
also known as N20 in the literature (\citealt{2015A&A...577A..48W}). At the
time of the core-collapse, the model is characterized by a $6 M_{\odot}$
helium core and a $10.3 M_{\odot}$ compact hydrogen envelope with a
radius of $3.4 \times 10^{12}$~cm ($\approx 49\,R_{\odot}$). In such a
way, the BSG model satisfies the helium core mass, hydrogen envelope
mass, luminosity and radius of Sk $-69^{\rm o}\, 202$ at collapse
(\citealt{1990ApJ...360..242S}).

The second model, B18.3, was proposed recently
(\citealt{2018MNRAS.473L.101U}) in the framework of the slow-merger
scenario (\citealt{2002MNRAS.334..819I}) and it denotes a $18.3
M_{\odot}$ BSG resulting from the merging of two massive stars with
$14 M_{\odot}$ and $9 M_{\odot}$, respectively. This model is
analogous to the merger model previously proposed by
\cite{2017MNRAS.469.4649M}. These models are particularly interesting
because they reproduce most of the observational constraints of Sk
$-69^{\rm o}\, 202$, such as the red-to-blue evolution, its lifetime
($\approx 20 000$~years), the total mass ($\approx 18.3 M_{\odot}$)
and position in the Hertzsprung-Russell diagram ($\log T_{\rm eff}
= 4.2$, $\log L/L_{\odot} = 4.9$) at collapse. Model B18.3 also
predicts a surface chemical composition before envelope deflation
of 0.139, 3.47 and 1.21 for He/H, N/C and N/O, respectively, in
agreement with observations. Model B18.3 describes an external
envelope which is quite similar to that of model N16.3 (see
Fig.~\ref{star_prog}). Otherwise, the internal structure of B18.3
differs significantly from that of N16.3. In particular, model B18.3
is characterized by a C+O core with a smaller size and with a profile
of $\rho r^3$ flatter than that of N16.3 (see Fig.~\ref{star_prog}).
We expect, therefore, that the SNRs from these two models differ
mainly for the structure and distribution of innermost ejecta rather
than for outermost ejecta.

In addition to the above models, we examined the model S19.8
(\citealt{2002RvMP...74.1015W, 2016ApJ...821...38S}) which denotes
a non-rotating, solar metallicity progenitor RSG with a main-sequence
mass $19.8 M_{\odot}$ which reduces to $15.9 M_{\odot}$ at collapse.
This model has a structure completely different from those of models
N16.3 and B18.3 (see Fig.~\ref{star_prog}). However it predicts
helium core mass, hydrogen envelope mass, and luminosity consistent
with those of Sk $-69^{\rm o}\, 202$ (although the stellar radius
is much larger than that of Sk $-69^{\rm o}\, 202$). The C+O layer
and the helium core mass are also similar to those of N16.3,
although the $\rho r^3$ gradient of the helium layer is very different
in the two models (\citealt{2016ApJ...821...38S}) (see
Fig~\ref{star_prog}). Model S19.8 was considered in order to
investigate how sensitive the remnant evolution is to the structure
of the progenitor star.

\subsection{Modeling the evolution of the supernova}
\label{SNmodel}

We followed the evolution of the core-collapse SN by adopting the
hydrodynamic model presented in \cite{2013ApJ...773..161O} but
extended to three dimensions. In Paper I, we describe in detail the
3D model and the corresponding numerical setup. In the following,
we summarize the main features of the model.

The SN evolution was modeled by numerically solving the time-dependent
hydrodynamic equations of mass, momentum, and energy conservation
in a 3D Cartesian coordinate system $(x, y, z)$. The model accounts
for the effects of gravity (both self-gravity and gravitational
effects of the central proto-neutron star), the fallback of
material on the proto-neutron star, the explosive nucleosynthesis
through a nuclear reaction network, the feedback of nuclear energy
generation, and the energy deposition due to radioactive decays of
isotopes synthesized in the SN explosion. The adopted equation of
state depends on the different physical regimes simulated (see Paper I):
at early phases, for $10^{-10}$~g~cm$^{-3} < \rho < 10^{11}$~g~cm$^{-3}$
and $10^4$~K$ < T < 10^{11}$~K, we adopted the Helmholtz equation of
state (\citealt{2000ApJS..126..501T}), which includes contributions from
radiation, fully ionized nuclei, and degenerate/relativistic electrons
and positrons; at later phases, we considered an equation of state that
includes contributions from the radiation and ideal gas of elements
(\citealt{2013ApJ...773..161O}) for $\rho < 10^{-8}$~g~cm$^{-3}$, and a smooth
transition between the latter equation of state and that of Helmholtz
for $10^{-8}$~g~cm$^{-3} < \rho < 10^{-7}$~g~cm$^{-3}$. The pressure
from radiation was neglected in the optically thin regime by adopting
an approximate approach described by \cite{2010ApJ...709...11J}.

The SN explosion is initiated by injecting kinetic and thermal energies
artificially around the composition interface of the iron core and
the silicon-rich layer. We modeled aspherical explosions with clumpy
structures to mimic the global anisotropies developing in magnetic
jet-driven SNe (\citealt{1999ApJ...524L.107K, 2003ApJ...598.1163M,
2009ApJ...696..953C, 2018MNRAS.478..682B}) or, alternatively, in
the context of neutrino heating mechanism, by convection in the
neutrino heating layers and the standing accretion shock instability
(SASI; e.g. \citealt{2006A&A...453..661K, 2009ApJ...694..664M,
2010PASJ...62L..49S, 2010ApJ...720..694N, 2012ApJ...749...98T,
2015A&A...577A..48W, 2016ARNPS..66..341J, 2017hsn..book.1095J,
2018PhRvD..98l3001W}). First we set the initial radial velocity as

\begin{equation}
v_{r} \propto r f(\theta)~,
\label{eq_shape}
\end{equation}

\noindent
where $f(\theta)$ is a function of $\theta$ in spherical coordinates
selected in order to produce a bipolar-like explosion in the light
of the observed elliptical shape of the inner ejecta distribution
in SN\,1987A (\citealt{2013ApJ...768...89L, 2016ApJ...833..147L}). More
specifically, we explored four cases (see Paper I for details): a
function with cosine, an exponential form, a power-law form, and
an elliptical form.  The degree of asymmetry of the explosion is
regulated by the parameter $\beta$ which is the ratio of the radial
velocity on the polar axis, $v_{\rm pol}$, to that on the equatorial
plane, $v_{\rm eq}$, at a given radius ($\beta = v_{\rm pol}/v_{\rm
eq}$). Then we imposed asymmetry across the equatorial plane by
changing the normalization of $v_r$ across this plane; this is
regulated by a parameter $\alpha = v_{\rm up}/v_{\rm dw}$, where
$v_{\rm up}$ and $v_{\rm dw}$ are the initial radial velocities at
a radius inside the shock at $\theta = 0^{\rm o}$ and $\theta =
180^{\rm o}$, respectively. This kind of asymmetry may be introduced
by the combined action of neutrino heating and SASI. Finally, we
introduced large amplitude perturbations ($30$\%) to the initial
radial velocities by superposing functions of $(\theta, \phi)$ in
spherical coordinates, based on real spherical harmonics (see Paper
I for details). We explored the space of parameters of the explosion
(energy and parameters regulating the initial asymmetry);
Table~\ref{sn_param} reports the range of parameters explored and
the parameters of the best-fit models N16.3, B18.3, and S19.8
(corresponding to models n16.3-high, b18.3-high, and s19.8-fid
respectively, in Table~2 of Paper I).

The simulations were performed using the adaptive mesh hydrodynamic code
FLASH (\citealt{for00}). The code consists of inter-operable modules
that can be combined to generate different astrophysical applications
and it was designed to make efficient use of massive parallel computers
using the message passing interface (MPI) library. For the 3D SN
simulations, we used the directionally split Eulerian version of the
piecewise parabolic method (PPM; \citealt{cole84}), which provides a
formal second-order accuracy in both space and time. Also a hybrid Riemann
solver which switches to an HLLE solver inside shocks was
used to avoid an odd-even instability (decoupling) that can develop
from shocks aligned with the grid (\citealt{quirk97}).

The gravitational effects are included by adopting a spherically symmetric
approximation\footnote{This approximation allowed us to reduce the
computational cost needed if, instead, the Poisson equation for self-gravity is
solved.}: first we calculated angle-averaged radial profiles of density
and, then, we estimated local gravitational potentials from enclosed
masses at each radius. A point source gravity from the mass of the
proto-neutron star was also included; the total mass that accretes onto
the proto-neutron star during the evolution was added to the point
mass (see Paper I for details).

The nuclear reaction network includes 19 species: neutron (n),
proton (p), $^{1}$H, $^{3}$He, $^{4}$He, $^{12}$C, $^{14}$N, $^{16}$O,
$^{20}$Ne, $^{24}$Mg, $^{28}$Si, $^{32}$S, $^{36}$Ar, $^{40}$Ca,
$^{44}$Ti, $^{48}$Cr, $^{52}$Fe, $^{54}$Fe, and $^{56}$Ni (see
\citealt{1978ApJ...225.1021W} for the network chain). The calculations
were performed by using the MA28 sparse matrix package (\citealt{duff86})
and a time integration scheme based on the Bader-Deuflhard method
(\citealt{bader83}), both available in FLASH. The distribution and
evolution of elements were followed by adopting a multiple fluids
approach. Each fluid is associated to one of the heavy elements of
the nuclear reaction network; the fluid evolution is calculated by
an advection equation which is solved in addition to the hydrodynamic
equations. The energy depositions due to radioactive decays of
$^{56}$Ni to $^{56}$Co to $^{56}$Fe are calculated by adopting the
method described by \cite{2009ApJ...693.1780J} (see also
\citealt{2013ApJ...773..161O} and Paper I for more details).

In order to follow the large physical scales spanned during the
remnant expansion, we followed a mesh strategy similar to that
proposed by \cite{2013ApJ...773..161O}. The initial computational
domain is a Cartesian box extending between $\approx -5000$~km and
$\approx 5000$~km in the $x$, $y$, and $z$ directions, with the SN
explosion assumed to sit at the origin of the coordinate system
$(x_0, y_0, z_0) = (0, 0, 0)$. In such a way, the domain includes
the inner regions of the oxygen-rich layer of the progenitor star
at collapse. In this mesh, we exploited the adaptive mesh refinement
(AMR) capabilities of FLASH (through the PARAMESH library;
\citealt{2000CoPhC.126..330M}) to reach a maximum spatial resolution
of $4.9$~km in low-resolution simulations and $2.4$~km in high-resolution
simulations. In Paper I, we describe in
detail the AMR strategy adopted for our SN simulations. Then the
computational domain was gradually extended as the forward shock
propagates: when the forward shock was close to one of the boundaries
of the Cartesian box, the physical size of the computational domain
was extended by a factor of 1.2 in all directions. The physical
quantities of the old domain were re-mapped in the new domain; the
physical quantities in the extended region were set to the values
of the pre-SN star and of the wind (after the shock breakout). This
approach is possible because the propagation
of the forward shock is supersonic and does not introduce errors
larger than 0.1\% after 40 re-mappings (\citealt{2013ApJ...773..161O}).
We found that about 75 re-mappings were necessary
to follow the propagation of the SN blast wave through the star
until the breakout of the shock wave at the stellar surface, about
20 hours after the core-collapse. The final domain extends between
$\approx -2\times 10^{14}$~cm and $\approx 2\times 10^{14}$~cm in
the $x$, $y$, and $z$ directions, leading to a maximum spatial
resolution of $\approx 2\times 10^{11}$~cm.

\subsection{Modeling the evolution of the supernova remnant}
\label{SNRmodel}

The output of the SN simulations about $20$ hours after the
core-collapse was used as initial condition for 3D MHD simulations
describing the interaction of the blast wave with the ambient medium.
The numerical setup for the SNR model has been described in detail
in \cite{2019A&A...622A..73O}. Briefly, the evolution and transition
from the SN phase to the SNR phase were modeled by numerically
solving the time-dependent MHD equations of mass, momentum, energy,
and magnetic flux conservation in a 3D Cartesian coordinate system
$(x,y,z)$.

The calculations were performed using PLUTO (\citealt{2007ApJS..170..228M,
2012ApJS..198....7M}), an advanced multi-dimensional Godunov-type
code for astrophysical plasmas and high Mach number flows. The code
was extended by additional computational modules to calculate the
deviations from temperature-equilibration between electrons and
ions, and the deviations from equilibrium of ionization of the most
abundant ions (see \citealt{2019A&A...622A..73O} for the details of
the implementation).
The former were implemented by including the almost instantaneous
heating of electrons at shock fronts up to $kT \approx 0.3$~keV by lower
hybrid waves (\citealt{2007ApJ...654L..69G}) and the effects of Coulomb
collisions for the calculation of ion and electron temperatures in
the post-shock plasma (\citealt{2015ApJ...810..168O}). The deviations
from equilibrium of ionization were calculated through the computation
of the maximum ionization age in each cell of the spatial domain
(\citealt{2015ApJ...810..168O}).

We followed the chemical evolution of the ejecta by adopting a
multiple fluids approach (\citealt{2016ApJ...822...22O}). Each fluid is
associated to one of the species modeled in the SN simulations and
initialized with the corresponding abundance in the output of the
SN simulations. As for the composition of the circumstellar medium
(CSM) around SN\,1987A, we adopted the abundances inferred from the
analysis of X-ray observations (\citealt{2009ApJ...692.1190Z}). This
allowed us to follow the chemical evolution of the ejecta and to
map the spatial distribution of heavy elements at different epochs
during the evolution of SN\,1987A. The different fluids mix together
during the remnant evolution: the density of a specific element in
a fluid cell at time $t$ is calculated as $\rho_{\rm el} = \rho
C_{\rm el}$, where $C_{\rm el}$ is the mass fraction of each element
and the index ``el'' refers to a different element.

We modeled the CSM as in previous works (\citealt{2015ApJ...810..168O,
2019A&A...622A..73O}). The immediate surrounding of the SN event
was modeled as a spherically symmetric wind characterized by a gas
density proportional to $r^{-2}$ (where $r$ is the radial distance
from the center of explosion), a mass-loss rate of $\dot{M}_{\rm
w} = 10^{-7} M_{\odot}$~year$^{-1}$, and a wind velocity $u_{\rm
w} = 500$~km~s$^{-1}$ (\citealt{2007Sci...315.1103M}). The termination
shock of the wind is located at $r_{\rm w} = 0.05$~pc. The
circumstellar nebula consists of an extended ionized H\,II region,
a dense inhomogeneous equatorial ring, and two less dense rings
lying in planes almost parallel to the equatorial one, but displaced
by about 0.4 pc above and below the central ring. The H\,II region
is uniform with density $n_{\rm HII}$; its inner edge in the
equatorial plane is at distance $r_{\rm HII}$ from the center of
explosion.  The central ring is composed of a uniform smooth component
and $N_{\rm cl}$ high-density spherical clumps mostly located in
its inner portion (\citealt{1995ApJ...452L..45C, 2005ApJS..159...60S}).
The uniform component has density $n_{\rm rg}$, radius $r_{\rm rg}$,
and an elliptical cross section with the major axis $w_{\rm rg}$
lying on the equatorial plane and height $h_{\rm rg}$. The high-density
clumps have a
diameter $w_{\rm cl}$ and their plasma density and radial distance
from SN\,1987A are randomly distributed around the values $<n_{\rm
cl}>$ and $<r_{\rm cl}>$ respectively. We explored the space of
parameters of the CSM around the fiducial values derived in a
previous work (\citealt{2015ApJ...810..168O}). Table~\ref{snr_param}
reports the range of parameters explored and the best-fit parameters
found for the three progenitor stars considered (N16.3, B18.3, and S19.8).

In order to trace the evolution of the different plasma
components (ejecta, H\,II region, and ring material), we introduced
passive tracers, each associated with a different component.
Each tracer is initialized to one in cells belonging to the plasma
component and to zero elsewhere. The continuity equations of the
tracers are solved in addition to our set of MHD equations. Some
tracers are also used to store information on the shocked plasma
(time, shock velocity, and shock position, i.e. Lagrangian coordinates,
when a cell of the mesh is shocked either by the forward or by the
reverse shock) which are required to synthesize the thermal X-ray
emission (see Sect.~\ref{sec:x-ray}). These last tracers are
initialized to zero.

The ambient magnetic field (most likely that originating from the
progenitor star) is modeled as in \cite{2019A&A...622A..73O} as a
``Parker spiral'', in analogy with the spiral-shaped magnetic field
on the interplanetary medium of the solar system
(\citealt{1958ApJ...128..664P}). This is the simplest field
configuration for a rotating magnetized progenitor star, resulting
from the rotation of the star and from the expanding stellar wind.
For our purpose, we considered a pre-SN magnetic field characterized
by an average strength at the stellar surface $B_0\approx 3$~kG
(well within the range inferred for magnetic massive stars, e.g.
\citealt{2009ARA&A..47..333D}) and a strength at the inner edge of
the nebula (at a distance from the center of explosion $r\approx
0.08$~pc) $B_1 \approx 1\,\mu$G. This pre-SN magnetic field
configuration leads to field strength within the nebula in agreement
with observations (\citealt{2018ApJ...861L...9Z}). As found in
\cite{2019A&A...622A..73O}, the field plays a role in preserving
inhomogeneities of the CSM (as the rings and the clumps) from
complete fragmentation by limiting the growth of hydrodynamic
instabilities that would develop at their boundaries
(\citealt{2008ApJ...678..274O}), whereas it does not influence the
overall expansion and evolution of the blast wave.

We adopted a mesh strategy similar to that used to model the
evolution of the SN explosion to follow the large physical scales
spanned during the remnant expansion (see Sect.~\ref{SNmodel} and
\citealt{2019A&A...622A..73O}). The initial computational domain
is a Cartesian box extending between $-3\times 10^{14}$~cm and
$3\times 10^{14}$~cm in the $x$, $y$, and $z$ directions, thus
including the final domain of the SN simulations (see Sect.~\ref{sec:x-ray}).
For the evolution of the SNR, the box was covered by a non-uniform
grid with the smallest mesh size (and the highest spatial resolution)
in the remnant interior. The grid is made of $1280^3$ zones and
consists of the following: i) a uniform grid patch with $512^3$
points, extending between $-7.5\times 10^{13}$~cm and $7.5\times
10^{13}$~cm in the $x$, $y$, and $z$ directions (including metal-rich
ejecta), and leading to a maximum resolution of $\approx 2.9\times
10^{11}$~cm, and ii) a non-uniform grid patch in the rest of the domain
with resolution ranging between $\approx 2.9\times 10^{11}$~cm close
to the high-resolution patch and $\approx 5.8\times 10^{11}$~cm
close to the border of the domain. This non-uniform mesh allows us
to describe in more detail the distribution of metal-rich ejecta
in the remnant interior. Then, as for SN simulations, we gradually
extended the computational domain by a factor 1.2 as the forward
shock propagates outward, and remapped the physical values to the
new domains.  All the physical quantities in the extended region
are set to the values of the pre-SN CSM. We found that about 49
re-mappings were necessary to follow the interaction of the blast
wave with the CSM during 50 years of evolution. The final domain
extends between $-1.3\times 10^{18}$~cm and $1.3\times 10^{18}$~cm
in the $x$, $y$, and $z$ directions, leading to a maximum spatial
resolution of $\approx 1.3\times 10^{15}$~cm in the remnant interior
and of $\approx 2.6\times 10^{15}$~cm in the outer layers of the
remnant. All physical
quantities were set to the values of the pre-SN CSM at all boundaries.

\subsection{Synthesis of X-ray emission}
\label{sec:x-ray}

From the model results, we synthesized the thermal X-ray emission
originating from the impact of the blast wave with the nebula, by
following the approach outlined in previous studies
(\citealt{2015ApJ...810..168O, 2019NatAs...3..236M}). The system
was rotated about the three axes to fit the orientation of the ring
in the plane of the sky inferred from the analysis of optical data
(\citealt{2005ApJS..159...60S}): $i_x = 41^{\rm o}$, $i_y = -8^{\rm
o}$, and $i_z = -9^{\rm o}$. Then, for each $j$-th cell of the
spatial domain, we derived: the maximum ionization age $\tau_{\rm
j} = n_{\rm ej} \Delta t_{\rm j}$ (where $ n_{\rm ej}$ is the
particle number density in the $j$-th domain cell and $\Delta t_{\rm
j}$ is the time since the plasma in the cell was shocked), the
emission measure em$_{\rm j} = n_{\rm Hj}^2 V_{\rm j}$ (where $n_{\rm
Hj}^2$ is the hydrogen number density in the cell and $V_{\rm j}$
is the cell volume), and the electron temperature $T_{\rm ej}$
(derived from the ion temperature, plasma density, and $\Delta
t_{\rm j}$ by assuming Coulomb collisions and starting from an
electron temperature at the shock front $kT = 0.3$~keV, which is
assumed to be the same at any time as a result of instantaneous
heating by lower hybrid waves; \citealt{2007ApJ...654L..69G,
2015ApJ...810..168O}). The X-ray emission was synthesized in the
$[0.1, 10]$~keV band from the values of $\tau_{\rm j}$, em$_{\rm
j}$, and $T_{\rm ej}$, assuming that the system is at a distance
$D = 51.4$ kpc (\citealt{1999IAUS..190..549P}), and by using the
non-equilibrium of ionization (NEI) emission model VPSHOCK available
in the XSPEC package along with the NEI atomic data from ATOMDB
(\citealt{2001ApJ...556L..91S}). For the CSM, we adopted the metal
abundances derived from the analysis of deep {\it Chandra} observations
of SN\,1987A (\citealt{2009ApJ...692.1190Z}); for the ejecta, we
used the metal abundances calculated during the ejecta evolution
after the core-collapse. Finally, the X-ray spectrum from each cell
was filtered through the photoelectric absorption by the interstellar
medium, assuming a column density $N_{\rm H} = 2.35\times 10^{21}$
cm$^{-2}$ (\citealt{2006ApJ...646.1001P}), and folded through the
instrumental response of either {\it XMM-Newton}/EPIC or {\it
Chandra}/ACIS.  X-ray lightcurves were derived by integrating the
spectra from the cells in spectral bands of interest and in the
whole spatial domain; X-ray images were derived by integrating the
spectra in bands of interest and along the line-of-sight (LoS).

\section{Results}
\label{results}

\subsection{Early distribution and mixing of ejecta in the remnant}
\label{sec_sn_evol}

We ran several 3D high-resolution simulations of the SN, searching
for the parameters (explosion energy, $E_{\rm exp}$, and parameters
for the initial asymmetry, $\alpha$ and $\beta$; see Sect.~\ref{SNmodel})
best reproducing the observed shifts and broadening of [Fe\,II]
lines (\citealt{1990ApJ...360..257H, 1994ApJ...427..874C,
1995A&A...295..129U}). Table~\ref{sn_param} reports the range of
parameters explored and the best-fit parameters for each pre-SN
models adopted.

\begin{figure}[!t]
  \begin{center}
  \leavevmode
  \epsfig{file=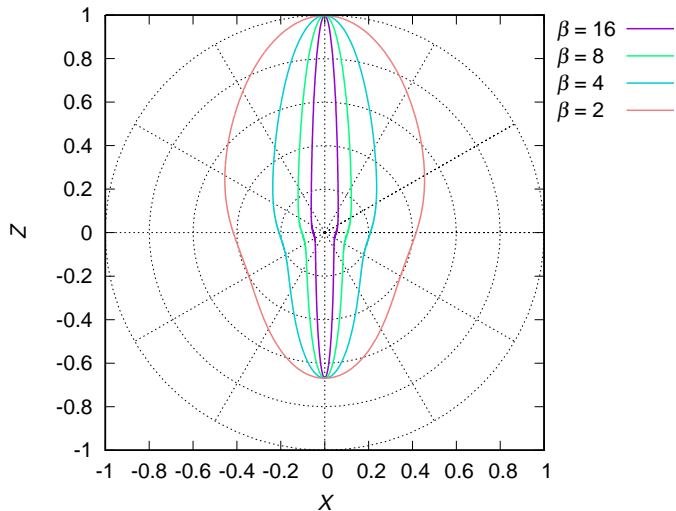, width=9.cm}
  \caption{Angle dependence of the initial radial velocity in the
  $[x, z]$ plane of the Cartesian coordinate, adopting an elliptical
  form for the function $f(\theta)$ in Eq.~\ref{eq_shape}. The
  different curves are for different values of $\beta = v_{\rm pol}/v_{\rm eq}$
  and $\alpha = v_{\rm up}/v_{\rm dw} = 1.5$.}
\label{heating}
\end{center}
\end{figure}

In Paper I, we describe in detail the evolution of the SN from the
energy release soon after the core-collapse to the breakout of the
shock wave at the stellar surface, covering about 20~hours of
evolution. In all the cases considered, the initial large
scale asymmetry leads to an expanding bipolar structure. During
the early phases, the bipolar structure is subject to development
of Kelvin-Helmholtz (KH) and Rayleigh-Taylor (RT) instabilities.
The growth of these instabilities depends on the degree of asymmetry
of the explosion: the higher the asymmetry, the faster their growth.
The morphology of the bipolar structure depends on the progenitor
star models: it is the narrowest in model B18.3 due to the flatter
$\rho r^3$ profile and smaller size of the C+O core (see
Fig.~\ref{star_prog}), which allow for a rapid expansion of the
shock wave. The initial bipolar structure is roughly kept up to the
shock breakout even if its morphology can be significantly modified
during the shock propagation due to structure of the stellar interior
and the growth of hydrodynamic instabilities. In Paper I, we found
that an elliptical form for the function $f(\theta)$ in Eq.~\ref{eq_shape}
is required to reproduce the bipolar-like explosion of SN\,1987A;
Fig.~\ref{heating} shows the corresponding angle dependence of the
initial radial velocity. Thus, in the present paper, we focused the
analysis only on this case.

\begin{table}
\caption{Explored range of parameters of the SN models, and parameters
of the best-fit models considered in this paper.}
\label{sn_param}
\begin{center}
\begin{tabular}{lcccc}
\hline
\hline
Parameter                           &  Range explored   &   N16.3  &  B18.3  &  S19.8  \\
\hline
$E_{\rm exp}$ [$10^{51}$~erg]       &  $[1.5-2.5]$      &   $1.93$ &  $2.01$ & $1.89$ \\
$\beta [= v_{\rm pol}/v_{\rm eq}]$  &  $[2-16]$         &   16     &  16     & 16     \\
$\alpha [= v_{\rm up}/v_{\rm dw}]$  &  $[1-2]$          &   1.5    &  1.5    & 1.5    \\
\hline
\end{tabular}
\end{center}
\end{table}

After the shock breakout, the blast wave expands through the wind
of the progenitor star before the impact onto the nebula. In this
phase, the metal-rich ejecta expand almost homologously, carrying
the fingerprints of the asymmetric explosion and of the structure
of the progenitor star. The X-ray emission arising from the shocked
wind and from the shocked ejecta is very faint due to the low values
of wind density. In fact, no significant X-ray emission has been
detected during the first two years of evolution of SN\,1987A.
Indeed, in this phase, the expanding ejecta have been detected in
the lines of [Fe\,II] 26 $\mu$m and 18 $\mu$m (e.g.
\citealt{1990ApJ...360..257H, 1994ApJ...427..874C, 1995A&A...295..129U}).
Thus, from the models, we synthesized these lines, taking into
account the Doppler shift due to bulk motion of material along the
LoS.

Following \cite{1990ApJ...360..257H}, we considered several
approximations and simplifications to synthesize the [Fe\,II] lines.
In particular, we assumed that the temperature of the emitting iron
is isothermal and that the level populations of iron depend only
on the temperature (\citealt{1990ApJ...360..257H}). In fact, our
models neglect the heating of ejecta due to the radioactive decay
of $^{56}$Co and their cooling due to optical emission lines, and
show small variations of the temperature of unshocked innermost
ejecta at the epoch of observations.  Nevertheless, even in the
presence of these heating and cooling mechanisms, our assumption is
reasonable because the heating and cooling rates scale similarly
with the iron density (\citealt{1988ApJ...329L..25C}).  We assumed
also that the lines of [Fe\,II] 26~$\mu$m and 18~$\mu$m are produced
by optically thin plasma. This assumption is supported by the
analysis of \cite{1990ApJ...360..257H} which shows that most of the
measured emission of [Fe\,II] lines comes from optically thin
regions. Another assumption is that the emission of [Fe\,II] lines
is proportional to the Fe density (\citealt{1990ApJ...360..257H}).
In the light of the above assumptions, we derived the line profiles
by simply calculating the distribution of iron mass vs. the LoS
velocity and, then, convolved the profiles with a Gaussian of size
$400$~km~s$^{-1}$, to approximate the resolution of the observed
spectra (\citealt{1990ApJ...360..257H}). Finally we explored different
explosion energies and asymmetries, and different orientations of
the asymmetry with respect to the LoS, to identify the physical and
geometrical properties of the SN explosion which led to the shifts
and broadening of [Fe\,II] lines in early observations (e.g.
\citealt{1990ApJ...360..257H, 1994ApJ...427..874C, 1995A&A...295..129U}).

We found that a high energy of the explosion is required to reproduce
the observations: at the shock breakout, our best-fit models have
an explosion energy of $\approx 2\times 10^{51}$~erg (see
Table~\ref{sn_param}). The explosion energy was constrained
from the comparison between the maximum iron velocities reached in
our models and those inferred from the analysis of [Fe\,II] lines
during the first two years of evolution, and from the comparison
between the X-ray lightcurves synthesized from the models and those
observed during the first 30 years of evolution (see
Sect.~\ref{x-ray_snr}).  Lower (higher) energies produce lower
(higher) iron velocities and lower (higher) X-ray fluxes at odds
with the observations. We note that the explosion energies found
are at the upper end of the range of values, $E_{\rm exp} = 0.8-2.0
\times 10^{51}$~erg, proposed in the literature for SN\,1987A (see
Table~1 in \citealt{2014ApJ...783..125H}). From bolometric lightcurve
fitting what can be constrained is not the explosion energy itself
but the ratio between the explosion energy, $E_{\rm exp}$, and the
mass of hydrogen envelope, $M_{\rm env}$ (e.g.
\citealt{1990ApJ...360..242S}). Thus, the estimates of explosion
energies from bolometric lightcurve fitting depend on the model
adopted for the progenitor star. Studies that considered progenitor
BSGs that evolved as single stars suggest explosion energies in the
range $1.2-1.5\times 10^{51}$~erg (e.g. \citealt{2019A&A...624A.116U}).
Studies that modeled explosions of BSGs from binary mergers (analogous
to model B18.3) found slightly higher values, $E_{\rm exp} \approx
1.7\times 10^{51}$~erg (\citealt{2019MNRAS.482..438M}). The difference
between merger models and single star models is basically due to
the fact that the helium core mass in the former is rather smaller
than that in the latter and that the envelope composition is
significantly different in the two cases.

The two parameters $\alpha$ and $\beta$ in Table~\ref{sn_param}
determine the level of asymmetry of the explosion and, therefore,
the shifts and broadening of [Fe\,II] lines. The $\alpha$ parameter
regulates the asymmetry across the equatorial plane so that its
main effect is to shift the centroid of the lines. The $\beta$
parameter is the ratio of the radial velocity on the polar axis to
that on the equatorial plane. In this case, the main effect is to
determine the broadening of Fe II lines. We started the exploration
of the parameter space with low values of $\beta$ and values of
$\alpha$ in the range $[1-2]$, and progressively increased $\beta$
up to when the observed line broadening was reproduced by one of
our models. We found that the observations require $\beta = 16$,
indicating a high degree of asymmetry in the explosion with the
radial velocity on the polar axis, $v_{\rm pol}$, 16 times larger
than the radial velocity on the equatorial plane, $v_{\rm eq}$, and
with the radial velocity at $\theta = 0^{\rm o}$, $v_{\rm up}$, 1.5
times larger than the radial velocity at $\theta = 180^{\rm o}$,
$v_{\rm dw}$. This kind of asymmetry may result from a strong
sloshing of the stalled shock in polar directions, due to the SASI
(e.g. \citealt{2003ApJ...584..971B}), or, alternatively, in magnetic
jet-driven SNe (e.g. \citealt{1999ApJ...524L.107K}), or in single-lobe
SNe (e.g. \citealt{2005ApJ...635..487H}). As a consequence, most
of the energy is released along the polar axis.

We note that, among the forms considered for the function
$f(\theta)$ in Eq.~\ref{eq_shape}, the elliptical is the only one
reproducing the observed line profiles for the range of $\beta$
explored (see Paper I for details). Values of $\beta> 16$ would
produce lines broader than observed. We cannot exclude that other
forms of $f(\theta)$ might reproduce the observed line profiles for
$\beta> 16$. This however would indicate an even more extreme
asymmetry in the explosion which seems to be unlikely. Finally, we
note that a higher (lower) explosion energy would require a lower
(higher) value of $\beta$. Explosion energies $> 2\times 10^{51}$~erg
would be out of the range of values inferred from the analysis of
observed bolometric lightcurves and, as mentioned before, values
of $\beta > 16$ are unlikely. Thus, the values selected (namely
$\beta = 16$ and $E_{\rm exp} \approx 2\times 10^{51}$~erg) for the
favorite models are a compromise between very energetic and highly
asymmetric explosions.

\begin{figure}[!t]
  \begin{center}
  \leavevmode
  \epsfig{file=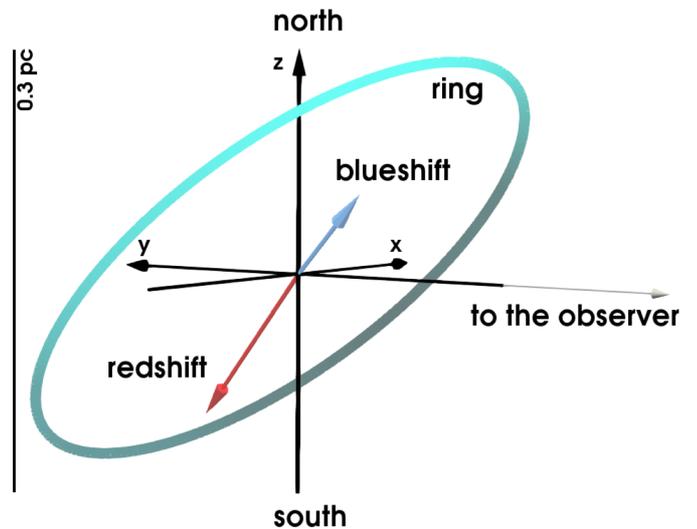, width=9.cm}
  \caption{Schematic view of the orientation of the initial asymmetry
  (lightblue and red arrows) with respect to the dense inner ring
  (lightblue circle) and to the LoS (gray arrow along negative
  $y$-axis). The negative $y$-axis points toward the observer. The
  yardstick indicating the length scale is on the left.}
\label{fig_orient}
\end{center}
\end{figure}

As for the orientation of the asymmetry with respect to the LoS,
we found that the polar axis should be almost lying in the plane
of the central ring of the nebula, roughly in the direction of the
LoS but with an offset of $40^{\rm o}$, and with the most energetic
lobe propagating away from the observer in the south (see
Fig.~\ref{fig_orient}). The projection of this axis in the plane
of the sky is offset by $15^{\rm o}$ from the northsouth axis.

\begin{figure*}[!t]
  \begin{center}
  \leavevmode
  \epsfig{file=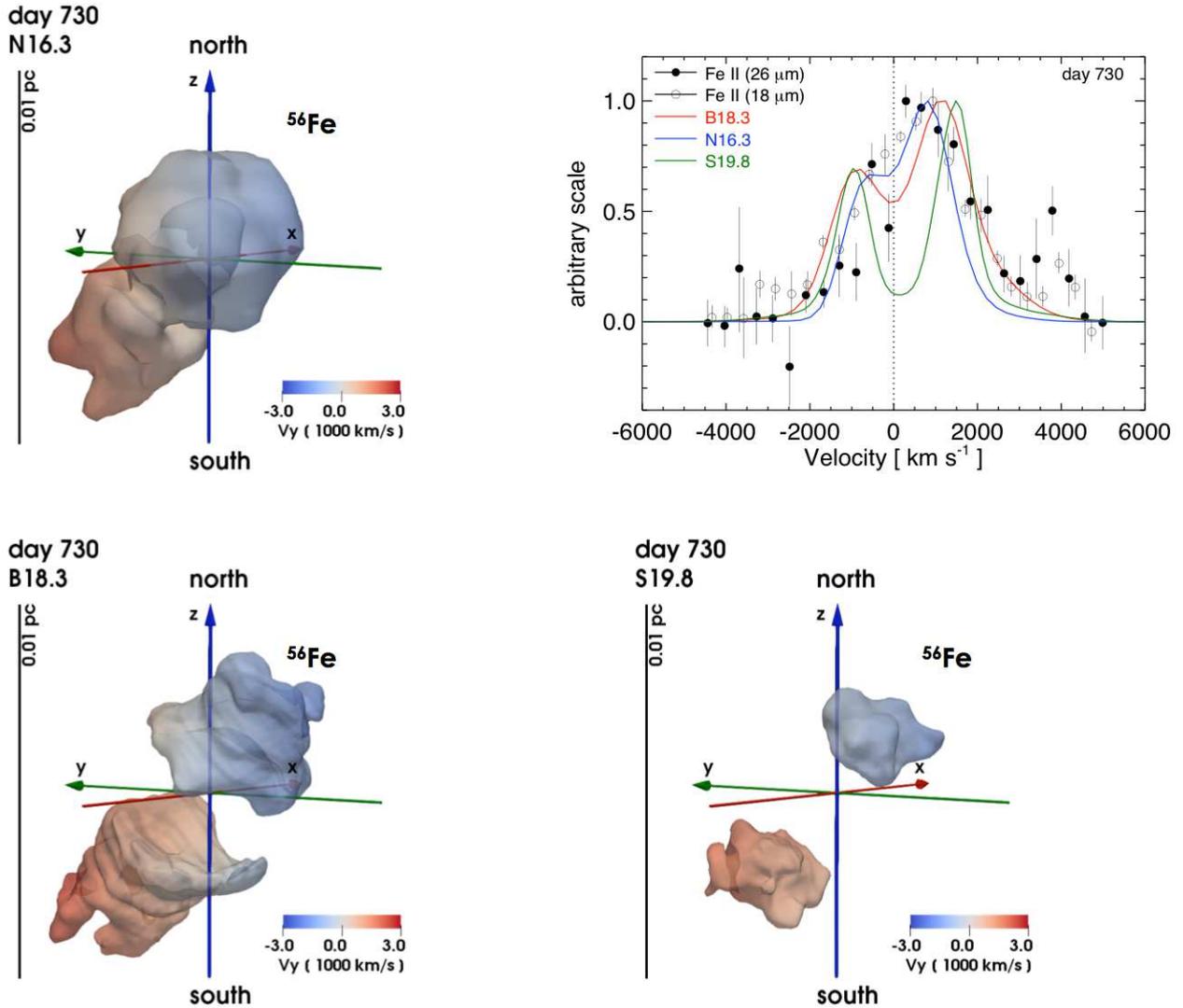, width=17.cm}
  \caption{Isosurfaces of the iron distribution for models N16.3
  (upper left panel), B18.3 (lower left), and S19.8 (lower right)
  at day 730. The semi-transparent isosurfaces correspond to a value
  of the iron density which is at 10\% of the peak density for each
  model (maximum iron density is $4.9\times 10^{-16}$~g~cm$^{-3}$
  for N16.3, $6.0\times 10^{-16}$~g~cm$^{-3}$ for B18.3, and
  $1.1\times 10^{-15}$~g~cm$^{-3}$ for S19.8). The
  colors give the velocity along the LoS in units of 1000~km~s$^{-1}$
  on the isosurface; the color coding is defined at the bottom of
  each panel. The yardstick indicating the length scale is on the
  left. The upper right panel shows the line profiles of [Fe\,II]
  26 $\mu$m and 18 $\mu$m of SN\,1987A observed during the first 2
  years of evolution (symbols; \citealt{1990ApJ...360..257H}) and
  synthesized from models N16.3, B18.3, and S19.8 at day 730 (solid
  lines). See online Movie 1 for an animation of these data; a
  navigable 3D graphic of the iron distribution for the three models
  is available at the link {https://skfb.ly/6P7sH}.}
\label{fig1}
\end{center}
\end{figure*}

As a result of the asymmetry, after the shock breakout, most of the
mass of $^{56}$Ni resides in two main clumps which propagate in
opposite directions with an angle of $40^{\rm o}$ from the LoS: a
massive one moves away from the observer in the south (red arrow
in Fig.~\ref{fig_orient}); the other (less massive) moves toward the
observer in the north (lightblue arrow in the figure). These clumps
may be those responsible for the so-called Bochum event observed
at days 20-100, when the H$\alpha$ profile exhibited two features
nearly symmetric on both the blue and red wings of the H$\alpha$
line (e.g. \citealt{1988MNRAS.234P..41H, 1989PASP..101..137P}). The
analysis of observations has suggested that these features are
probably due to two $^{56}$Ni clumps ejected during the explosion
(e.g. \citealt{1995A&A...295..129U, 1997ApJ...486.1026N,
2000ApJS..127..141N}) and propagating in opposite directions with
an angle ranging between $31^{\rm o}$ and $45^{\rm o}$ from the LoS
(e.g.  \citealt{1995A&A...295..129U, 2002ApJ...579..671W}). Our
modeled $^{56}$Ni clumps fit well in this scenario.

\begin{figure*}[!t]
  \begin{center}
  \leavevmode
  \epsfig{file=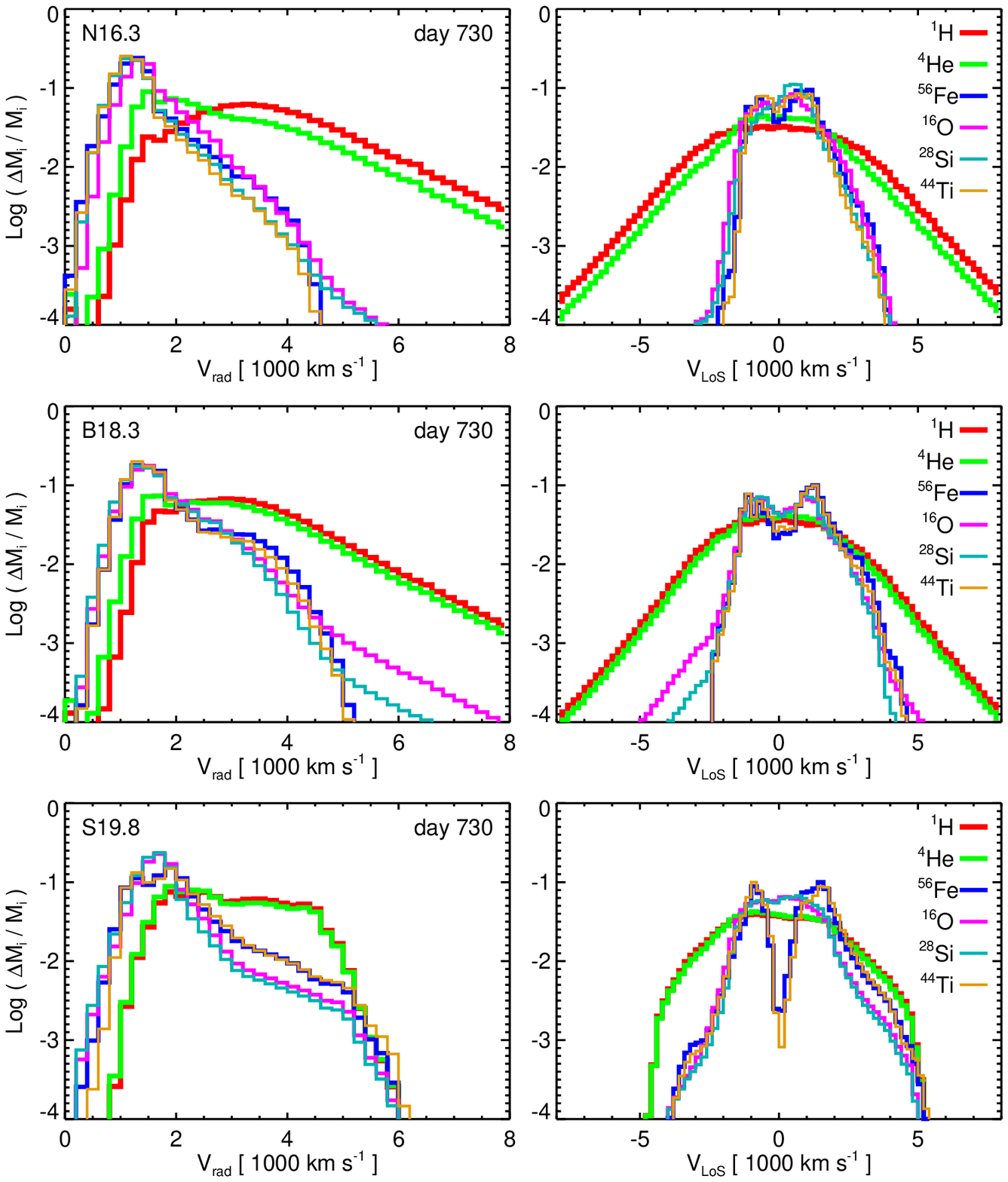, width=13.cm}
  \caption{Mass distributions of $^1$H, $^4$He, $^{56}$Ni, $^{16}$O,
  $^{28}$Si, and $^{44}$Ti versus radial velocity $v_{\rm rad}$ (left
  panels) and velocity component along the LoS,
  $v_{\rm LoS}$ (right panels) at day 730 for models N16.3 (top), B18.3
  (middle), and S19.8 (bottom).}
\label{vel_prof}
\end{center}
\end{figure*}

The distribution of iron-rich ejecta at day 730 (the time of the
first output of SNR simulations), after the decay of $^{56}$Ni in
$^{56}$Fe, reflects the initial large-scale asymmetry (see
Fig.~\ref{fig1}; see also online Movie 1). The figure also compares
the line profiles of [Fe\,II] 26 $\mu$m and 18 $\mu$m synthesized
from the models with those observed (\citealt{1990ApJ...360..257H}).
The only model able to reproduce the observed redshift and broadening
of [Fe\,II] lines and, in particular, the wings of the profiles
extending up to velocities $\approx 3000$~km~s$^{-1}$, is B18.3,
although it underestimates the mass of iron with low velocity (the
dip at the center of the line) and does not reproduce the feature
at $v\approx 4000$~km~s$^{-1}$, due to a fast isolated clump of
iron not modeled by our simulations. This suggests that the initial
large-scale asymmetry was more complex than modeled here. The other
models do not account for iron-rich ejecta with velocities $>
2200$~km~s$^{-1}$, for all the explosion energies and asymmetries
explored (see Table~\ref{sn_param}).

The differences in the line profiles synthesized from the models
can be further investigated by deriving the mass distributions of
selected elements versus the radial velocity, $v_{\rm rad}$, at day
730. The left panels of Fig.~\ref{vel_prof} shows these distributions
for models N16.3, B18.3, and S19.8. In the figure, $\Delta M_i$ is
the mass of the $i$-th element in the velocity range $[v - v +
\Delta v]$, and $M_i$ is the total mass of the $i$-th element. The
velocity is binned with $\Delta v = 200$~km~s$^{-1}$. At day 730,
we assumed that all $^{56}$Ni has already decayed in $^{56}$Fe; thus
the figure reports only the distribution of the latter. The figure
shows that the metal distribution can be strongly affected by the
internal structure (density and temperature) of the progenitor star.
In particular, the model adopting a progenitor RSG (S19.8) shows
strong differences with respect to the other two cases which adopt
a progenitor BSG (N16.3 and B18.3). This is a direct consequence
of the distinct structure of the progenitor stars (see
Fig.~\ref{star_prog}).  In model S19.8, the various species (and
in particular $^{1}$H and $^{4}$He) cannot reach velocities larger
than $\approx 6000$~km~s$^{-1}$ at variance with the other two
models. This is mainly due to the structure of the extended hydrogen
envelope of the RSG; in this case the blast wave is continuously
decelerated during its propagation through the hydrogen envelope
(see Paper I). As a consequence, the velocity of the forward shock
after the breakout at the stellar surface is significantly lower
in model S19.8 than in models N16.3 and B18.3, and $\approx 60$\%
of the total explosion energy in S19.8 is kinetic (whilst, in the
other two models, more than $90$\% of the explosion energy is
kinetic).

\begin{table*}
\caption{Explored and best-fit parameters of the CSM for the models of SN\,1987A.}
\label{snr_param}
\begin{center}
\begin{tabular}{lclcccc}
\hline
\hline
CSM       & Parameters   & Units & Range    &  N16.3  & B18.3 & S19.8\\
component &              &       & explored &    & \\
\hline
BSG wind: & $\dot{M}_{\rm w}$  & ($M_\odot$~yr$^{-1}$) &  $10^{-7}$ & $10^{-7}$ & $10^{-7}$ & $10^{-7}$   \\
          & $v_{\rm w}$        & (km~s$^{-1}$)   &  500          & 500   & 500  & 500\\
          & $r_{\rm w}$        & (pc)            &  0.05         & 0.05  & 0.05 & 0.05 \\
\hline
H\,II region: & $n_{\rm HII}$  &  (cm$^{-3}$)  &  $[10-100]$    & 50     & 50    & 50 \\
              & $r_{\rm HII}$  &  (pc)         &  $[0.08-0.2]$  & 0.08   & 0.08  & 0.08 \\
\hline
Ring:   & $n_{\rm rg}$   &  ($10^3$ cm$^{-3}$) &  1     & 1  & 1 & 1 \\
        & $r_{\rm rg}$   &  (pc)               &  $[0.15-0.25]$  & 0.2  & 0.19 & 0.19  \\
        & $w_{\rm rg}$   &  ($10^{17}$ cm)     &  $[1.7-2.5]$    & 2.35 & 2.35 & 2.35  \\
        & $h_{\rm rg}$   &  ($10^{16}$ cm)     &  $[2-4]$        & 2    & 2    & 2 \\
\hline
Clumps: & $<n_{\rm cl}>$ &  ($10^4$ cm$^{-3}$) &  $[2-3]$        & 3     & 3    & 3    \\
        & $<r_{\rm cl}>$ &  (pc)               &  $[0.15-0.25]$  & 0.21  & 0.2  & 0.2  \\
        & $w_{\rm cl}$   &  ($10^{16}$ cm)     &  $[2-4]$        & 2.5   & 2.5  & 2.5 \\
        & $N_{\rm cl}$   &                     &  $[40-60]$      & 50    & 50   & 50   \\
\hline
\end{tabular}
\end{center}
\end{table*}

In the case of the two progenitor BSGs (N16.3 and B18.3), although
their distributions are similar for all nuclear species, some evident
differences appear at the high-velocity tail of the $^{56}$Fe,
$^{16}$O, $^{28}$Si, and $^{44}$Ti distributions. In particular,
in model B18.3, these species can reach higher velocities and they
have more mass in the high-velocity tail of the distributions than
in model N16.3. For instance, the fastest $^{56}$Fe (product of
decay from $^{56}$Ni) mixes up to velocities of $\approx 5000$~km~s$^{-1}$
in model B18.3, while it is about 10\% slower in model N16.3. This
is basically due to the fact that the velocity of the expanding
nickel and, then, of its product of decay, iron, depends on the
structure of the overlying C+O and He layers (see the gradients of
$\rho r^3$ radial profiles in Fig.~\ref{star_prog}).  High velocities
can be obtained if the nickel is able to reach the hydrogen layer
before the development of a strong reverse shock during the propagation
of the blast wave through the He layer. In model B18.3, C+O and He
layers are less extended and less dense than in model N16.3 (see
Fig.~\ref{star_prog}). In this way, the nickel can penetrate
efficiently through the helium and hydrogen layers. As a result,
the metal-rich ejecta in model B18.3 can expand faster than in model
N16.3, resulting in a larger matter mixing.  Some differences between
models N16.3 and B18.3 are evident also in the high velocity tails
of $^{1}$H and $^{4}$He where, in N16.3, these species have higher
velocities and more mass than in B18.3.  One of the consequences
is that the shock after the breakout at the stellar surface is
slightly faster in model N16.3.

The lines of [Fe\,II] 26 $\mu$m and 18 $\mu$m (shown in Fig.~\ref{fig1})
are expected to reflect the differences in the mass distributions
versus the velocity component along the LoS, $v_{\rm LoS}$, through
different Doppler shifts and broadenings. The right panels in
Fig.~\ref{vel_prof} shows the mass distributions versus $v_{\rm
LoS}$, assuming the orientation of the initial asymmetry required
to fit the observations (see Fig.~\ref{fig_orient}). The differences
noted in the distributions versus $v_{\rm rad}$ are also evident
in the distributions versus $v_{\rm LoS}$. As a consequence, the
line profiles reflect the distinct structure of the corresponding
progenitor stars and, besides to constrain the explosion energy and
asymmetry, enabled us to identify the progenitor star most appropriate
for SN\,1987A. The comparison of simulations with observations of
SN\,1987A (e.g. \citealt{1990ApJ...360..257H}) showed that the model
best reproducing the observations is B18.3 (see Fig.~\ref{fig1}).

\subsection{Evolution of the X-ray emitting remnant}
\label{x-ray_snr}

\begin{figure*}[!t]
  \begin{center}
  \leavevmode
  \epsfig{file=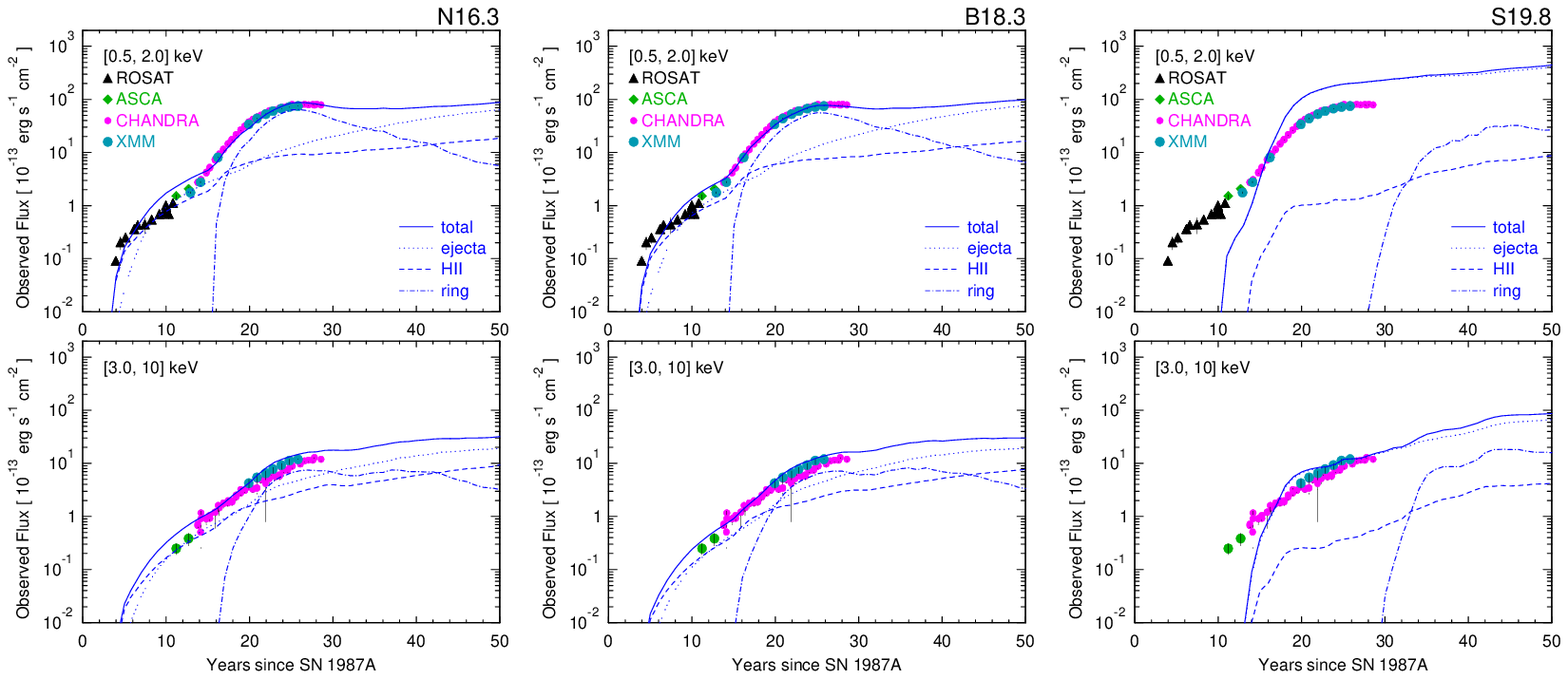, width=18.cm}
  \caption{X-ray lightcurves (solid lines) in the $[0.5, 2]$~keV
  (upper panels) and $[3, 10]$~keV (lower panels) bands synthesized
  from models N16.3 (left panels), B18.3 (center panels), and S19.8
  (right panels) compared to the lightcurves of SN\,1987A observed
  with Rosat (black triangles; \citealt{2006A&A...460..811H}), \emph{ASCA}
  (green diamonds; \citealt{2015ApJ...810..168O}), \emph{Chandra}
  (magenta circles; \citealt{2013ApJ...764...11H, 2016ApJ...829...40F})
  and \emph{XMM-Newton} (cyan circles; \citealt{2006A&A...460..811H,
  2012A&A...548L...3M}). Dotted, dashed, and
  dot-dashed lines indicate the contribution to emission from the
  shocked ejecta, the shocked plasma from the H\,II region, and the
  shocked plasma from the ring, respectively.}
\label{lc_xray}
\end{center}
\end{figure*}

We evolved the models for 50 years, describing the interaction of
the blast wave with the nebula. We ran several 3D high-resolution
simulations of the SNR to explore the model parameter space,
synthesizing (remote sensing) observables to compare the model
results with available observations (see section~\ref{sec:x-ray}).
In particular, we explored several density structures of the nebula
within and around the range of parameters discussed in the literature
(e.g. \citealt{2005ApJS..159...60S}) , searching for a set of values
that best reproduces the X-ray lightcurves and morphology of SN\,1987A
as observed in the last 30 years (e.g. \citealt{2015ApJ...810..168O}).
The range of parameters explored and the best-fit parameters found
for the three models (N16.3, B18.3, and S19.8) are shown in
Table~\ref{snr_param}.

\begin{figure*}[!t]
  \begin{center}
  \leavevmode
  \epsfig{file=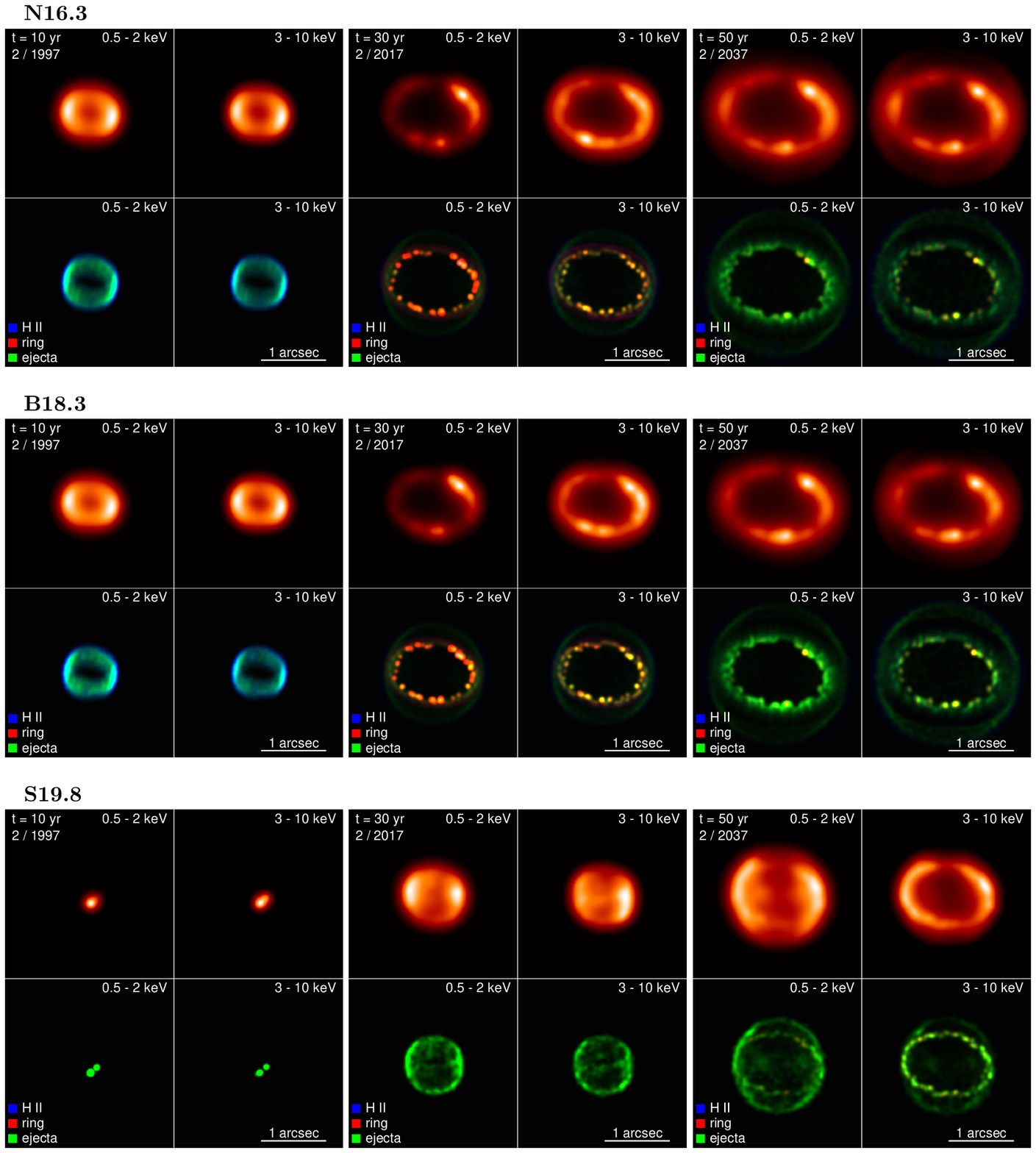, width=17.cm}
  \caption{Synthetic X-ray emission maps from models N16.3 (top),
  B18.3 (middle), and S19.8 (bottom) at the labeled times. Each
  panel consists of 4 quadrants: emission maps in the $[0.5,2]$~keV
  (upper left) and $[3,10]$~keV (upper right) bands, and corresponding
  three-color composite representations (lower left and right).
  Each image is normalized to its maximum for visibility; the X-ray
  maps were convolved with a Gaussian of size 0.15 arcsec to
  approximate the spatial resolution of {\it Chandra}; the three-color
  composite images were smoothed with a Gaussian of 0.025 arcsec;
  the colors in the composite show the emission from shocked ejecta
  (green), shocked plasma from the ring (red), and from the H\,II
  region (blue). See online Movies 2, 3, and 4
  for animations of these data.}
\label{xray_maps}
\end{center}
\end{figure*}

We found that the remnant evolution is similar in models which
consider a progenitor BSG (N16.3 and B18.3), whilst the evolution
in the case of a progenitor RSG (model S19.8) largely differs from
the others. The remnant hits the inhomogeneous nebula around year
3 in models N16.3 and B18.3 and around year 11 in model S19.8. In
the latter case, the forward shock is slower than in the other
models due to the early deceleration suffered by the shock during
its propagation through the extended hydrogen envelope of the RSG
(see section~\ref{sec_sn_evol}). The impact of the blast wave onto
the nebula produces a sudden increase of X-ray flux either due to
significant X-ray emission arising from the shocked gas of the H\,II
region (in models N16.3 and B18.3) or because of large X-ray emission
produced by the shocked outermost ejecta (in model S19.8).
Fig.~\ref{lc_xray} shows the X-ray lightcurves synthesized from the
three models. Online Movies 2, 3, and 4 show the evolution of the
3D volume rendering of particle density of the shocked
plasma (left panel), and the corresponding synthetic X-ray emission
maps in the $[0.5, 2]$~keV and $[3, 10]$~keV bands (upper right
panels), and three-color composite representations (lower right
panels), for models N16.3, B18.3, and S19.8, respectively; the
synthetic X-ray emission maps are also shown in Fig.~\ref{xray_maps}.

During the remnant-nebula interaction, the evolution follows three
phases as also found in previous studies (\citealt{2015ApJ...810..168O,
2019A&A...622A..73O}): in the first phase, the blast wave travels
through the H\,II region; in the second, it interacts with the dense
equatorial ring; finally, in the third phase, it leaves the ring
and starts to travel through a less dense environment. In models
N16.3 and B18.3, the first phase occurs between years 3 and 15 when
the X-ray emission is dominated by shocked material from the H\,II
region and from the outermost ejecta (see also upper and middle
panels in Fig.~\ref{xray_maps} and online Movies 2, and 3). In model
S19.8, this phase lasts between year 11 (when the blast hits the
H\,II region) and year 22 (when the blast reaches the ring) and,
at variance with the other cases, the X-ray emission is always
largely dominated by shocked ejecta (see right panels of
Fig.~\ref{lc_xray}, lower panels of Fig.~\ref{xray_maps}, and online
Movie 4).

The remnant enters the second phase when the blast wave hits the
dense equatorial ring. This occurs around year 15 in models N16.3
and B18.3 and around year 22 in model S19.8. In this phase the
forward shock travels through the ring, and the X-ray emission from
the shocked material of the ring becomes the dominant component in
models N16.3 and B18.3 (see Figs.~\ref{lc_xray} and \ref{xray_maps},
and online Movies 2, and 3). Conversely, the X-ray emission continues
to be largely dominated by shocked ejecta in model S19.8 (see
Figs.~\ref{lc_xray} and \ref{xray_maps}, and online Movie 4) and
this makes this model totally different from the others. In this
phase, the ambient magnetic field preserves the ring from erosion
by hydrodynamic instabilities and the ring survives the passage of
the blast (\citealt{2019A&A...622A..73O}). From year 26, the ring
is fully shocked in models N16.3 and B18.3 and its X-ray emission
gradually fades away. In model S19.8, the ring is fully shocked
around year 42.

During the whole evolution, the reverse shock travels through the
inner envelope of the SNR and the X-ray emission from shocked ejecta
gradually increases as the shock encounters higher and higher
densities of ejecta. In models N16.3 and B18.3, the third phase of
evolution begins around year 34, when the blast wave, after leaving
the ring, travels through a less dense environment and the X-ray
emission becomes dominated by shocked ejecta (Figs.~\ref{lc_xray} and
\ref{xray_maps}, and online Movies 2, and 3). In model S19.8, the
blast leaves the ring around year 42, but the emission continues
to be dominated by shocked ejecta (see online Movie 4).

By comparing the synthetic X-ray lightcurves with those observed,
the models best representing the X-ray observations are those
considering a progenitor BSG (N16.3 and B18.3). In the case of model
S19.8 (with a progenitor RSG), we did not find a set of parameters
of the CSM compatible with that inferred from observations
(e.g. \citealt{2005ApJS..159...60S}) which is able to reproduce the observed
X-ray lightcurves; this is mainly due to the fact that, in this
model, the X-ray emission is largely dominated by shocked ejecta
and this flux contribution is well above the observed values during
the whole interaction of the remnant with the nebula. The synthetic
lightcurves in this case are almost insensitive to the parameters
of the CSM. Clearly the X-ray emitting remnant of SN\,1987A keeps
memory of the structure of the progenitor star. As for models N16.3
and B18.3, both are able to reproduce in detail the observed
lightcurves with similar parameters of the CSM (compatible with
those inferred from observations; e.g. \citealt{2005ApJS..159...60S}). In
fact, the structure of the outermost ejecta (namely those interacting
with the nebula in the period analyzed) is similar in these two
models (see Fig.~\ref{star_prog}).

\begin{figure}[!t]
  \begin{center}
  \leavevmode
  \epsfig{file=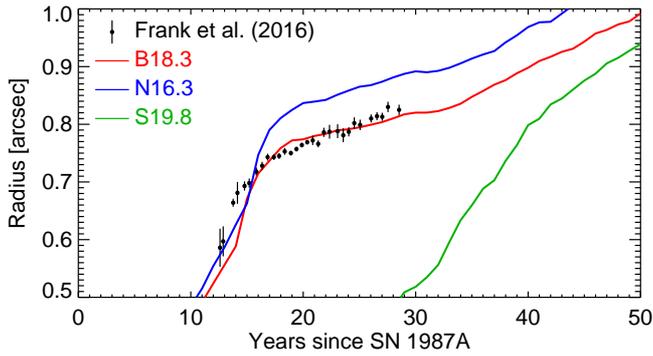, width=9.cm}
  \caption{Best-fit radius of the X-ray emitting torus vs time
  inferred from the analysis of actual X-ray observations
  (symbols; \citealt{2016ApJ...829...40F}) and that derived from the
  analysis of synthetic X-ray images for models B18.3 (red solid
  line), N16.3 (blue), and S19.8 (green).}
  \label{radius}
\end{center}
\end{figure}

To further investigate which model between N16.3 and B18.3 better
fits the X-ray observations, we compared the size of the expanding
X-ray emitting torus visible in synthetic images (see Fig.~\ref{xray_maps})
with the size observed in actual X-ray observations. To this end,
we analyzed the synthetic X-ray images in the $[0.5-10]$~keV band,
adopting a methodology analogous to that used for actual X-ray
observations (\citealt{2016ApJ...829...40F}), and assuming a distance
$D = 51.4$~kpc from SN\,1987A (\citealt{1999IAUS..190..549P}), the
same used for the synthesis of X-ray emission. Fig.~\ref{radius}
shows the best-fit radius of the torus vs time inferred from the
analysis of actual observations (symbols; \citealt{2016ApJ...829...40F})
and those derived from the synthetic images (solid lines). The
observed torus radii at different epochs are nicely reproduced by
model B18.3 which is able to describe the rapid expansion of the
torus between years 10 and 15 (when the blast was propagating through
the H\,II region) and the slower expansion of the torus between
years 15 and 30 (when the blast was propagating through the denser
material of the equatorial ring). Then, the model predicts a faster
expansion of the torus after year 32, namely after the blast has
left the ring and started to travel through a less dense environment.

The other two models fail to reproduce the observed expansion of
the X-ray emitting torus: model N16.3 roughly reproduces the expansion
when the blast wave was traveling through the H\,II region, but it
overestimates the torus radius after year 15, namely when the blast
wave was traveling through the ring; model S19.8 does not reproduce
the observed expansion of the torus and largely underestimates its
radius during the whole evolution. In fact, model N16.3 requires a
size of the equatorial ring slightly larger than in B18.3 to fit
the X-ray lightcurves (see Table~\ref{snr_param} and Fig.~\ref{lc_xray});
this is necessary because, in N16.3, the forward shock after the
breakout at the stellar surface is faster than in B18.3 (due to the
structure of the progenitor star; see Sect.~\ref{sec_sn_evol}).
As a consequence, the radius of the X-ray emitting torus in N16.3
is overestimated when the blast is traveling through the equatorial
ring. As it concerns model S19.8, it largely underestimates the
torus size during the whole evolution due to a significant lower
velocity of the forward shock caused by the early deceleration
suffered by the shock during its propagation through the extended
hydrogen envelope of the progenitor RSG (see Sect.~\ref{sec_sn_evol}).

We note that a smaller ring size in model N16.3 (necessary to fit
the observed radii of the torus) would have produced a steepening
of the soft X-ray lightcurve well before that observed when the
blast hits the ring. In other words, model N16.3 can reproduce
either the X-ray lightcurves or the size of the X-ray emitting
torus, whilst model B18.3 is able to naturally reproduce both at the
same time.

\subsection{Distribution of $^{44}$Ti in the evolved remnant}

During the interaction of the blast wave with the inhomogeneous
nebula, the metal-rich ejecta in the remnant interior expand and
form a bipolar structure which lies almost in the plane of the ring
and which reflects the initial large-scale asymmetry (see
Fig.~\ref{fig_orient}). The early $^{56}$Ni clumps and the distribution
of $^{56}$Fe (Fig.~\ref{fig1}), all lie along this structure. The
projection of the bipolar structure in the plane of the sky is
offset by $15^{\rm o}$ from the northsouth axis and coincides with
an elongated region, evident in maps of [Si\,I]+[Fe\,II] lines of
near-infrared and optical observations of SN\,1987A
(\citealt{2016ApJ...833..147L}), that lies roughly in the direction
of the northsouth axis but with an offset by $10^{\rm o}-20^{\rm
o}$. The ejecta probed by these emission lines are dominated by
radioactive input from $^{44}$Ti decay.

\begin{figure*}[!t]
  \begin{center}
  \leavevmode
  \epsfig{file=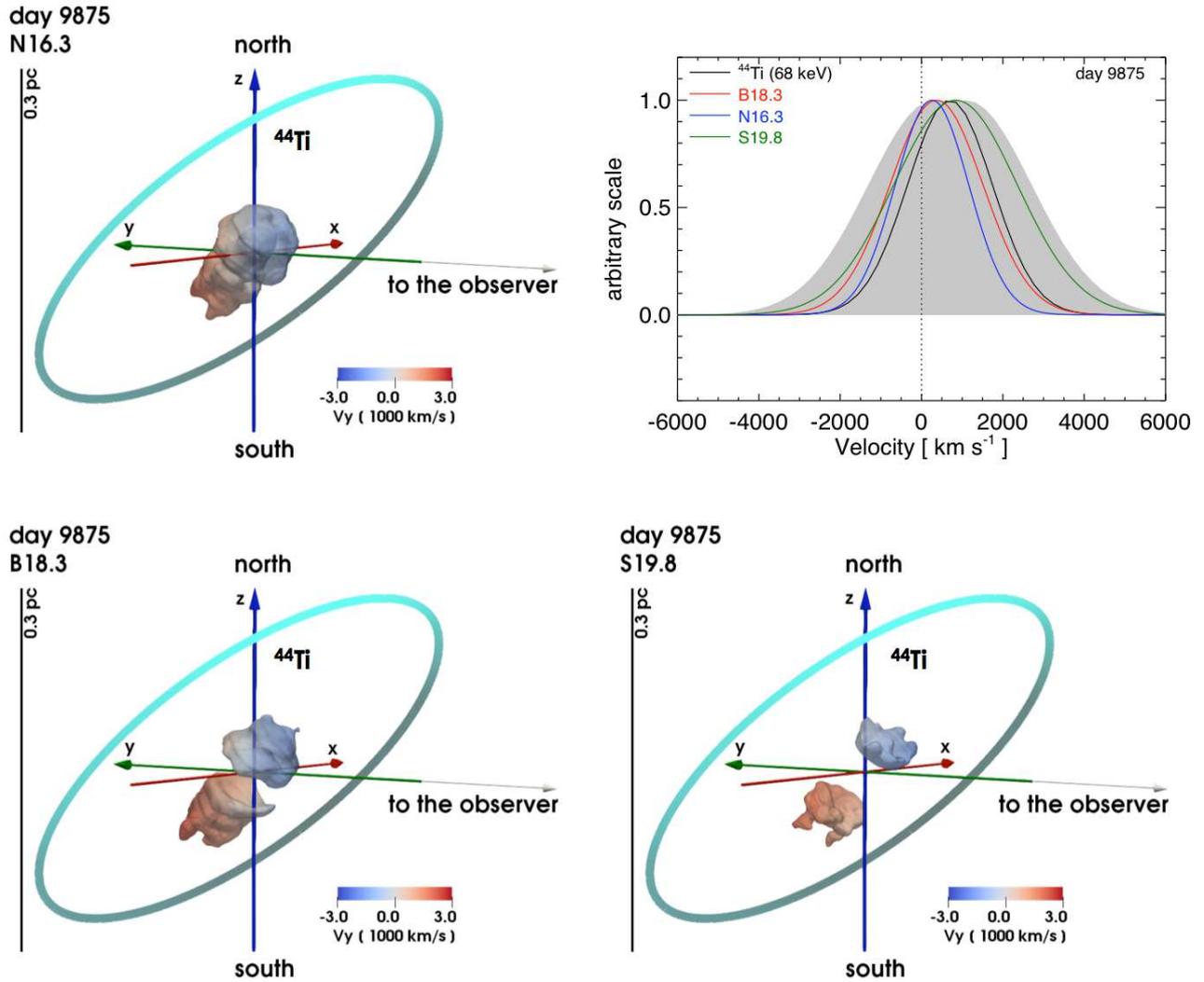, width=17.cm}
  \caption{Isosurfaces of the titanium
  distribution for models N16.3 (upper left), B18.3 (lower left),
  and S19.8 (lower right) at day 9875. The semi-transparent isosurface
  corresponds to a value of the titanium density which is at 10\%
  of the peak (maximum titanium density is $8.6\times 10^{-22}$~g~cm$^{-3}$
  for N16.3, $6.9\times 10^{-22}$~g~cm$^{-3}$ for B18.3, and
  $2.3\times 10^{-21}$~g~cm$^{-3}$ for S19.8). The colors give the
  velocity along the LoS in units of 1000 km~s$^{-1}$ on the
  isosurface; the color coding is defined at the bottom of each
  panel. The ring indicates the position of the dense inner ring.
  The yardstick on the left indicates the length scale. The upper
  right panel shows the best-fit Gaussian for the $^{44}$Ti lines
  observed between 2012 and 2014 (black line;
  \citealt{2015Sci...348..670B}) and lines profile synthesized from
  models N16.3 (blue line), B18.3 (red line), and S19.8 (green line)
  at day 9875. The shaded region represents the 90\% confidence
  area for the position of the peak and for the width of the Gaussian.
  See online Movie 5 for an animation of these data; a navigable
  3D graphic of the titanium distribution for the three models is
  available at the link {https://skfb.ly/6P7sI}.}
\label{fig_Ti}
\end{center}
\end{figure*}

Fig.~\ref{fig_Ti} and the online Movie 5 show the spatial distribution
of $^{44}$Ti for the three models around day $\approx 10000$, when
NuSTAR observations have resolved the emission lines from decay of
$^{44}$Ti at $67.87$ and $78.32$~keV (\citealt{2015Sci...348..670B}).
As found for $^{56}$Ni and $^{56}$Fe, again most of the mass of
$^{44}$Ti resides in two clumps moving in opposite directions along
the bipolar structure (Fig.~\ref{fig_Ti}, and online Movie 5):
the most massive clump propagates away from the observer in the
south at an angle of $40^{\rm o}$ from the LoS.  Thus the ejecta
powered by radiative decay of $^{44}$Ti are expected to lie
preferentially along the bipolar structure, consistently with the
observed maps of [Si\,I]+[Fe\,II] lines (\citealt{2016ApJ...833..147L}).

The evident differences in the distributions of the models originate
again from the distinct structure of the corresponding progenitor
stars. In model N16.3, most of the mass of $^{44}$Ti is concentrated
in a single clump elongated along the bipolar structure. The portion
of the clump expanding away from the observer includes $\approx
61$\% of the total mass of $^{44}$Ti (i.e. $\approx 5.4\times
10^{-4}\,M_{\odot}$). In the other two models (B18.3 and S19.8),
most of the mass resides in two clumps moving in opposite directions
along the bipolar structure. In both cases, the most massive clump
propagates away from the observer in the south with $\approx 64$\%
of the titanium mass (total mass $\approx 5.7\times 10^{-4}\,M_{\odot}$)
in B18.3, and $\approx 60$\% of the titanium mass (total mass
$\approx 8.9\times 10^{-4}\,M_{\odot}$) in S19.8. Considering that
the half life of $^{44}$Ti is 60 years, the total $^{44}$Ti mass
decreases by about 27\% after $\approx 10000$~days, leading to
$\approx 3.9\times 10^{-4}\,M_{\odot}$ for N16.3, $\approx 4.2\times
10^{-4}\,M_{\odot}$ for B18.3, and $\approx 6.5\times 10^{-4}\,M_{\odot}$
for S19.8.

We note that all the models overestimate the total mass of titanium
inferred from observations which, in general, fall in the range
$[0.5 - 2]\times 10^{-4}\,M_{\odot}$ (\citealt{2014ApJ...792...10S,
2015Sci...348..670B}). This is basically due to the limited number
of nuclei included in the adopted nuclear reaction network. In fact,
the mass of titanium can be overestimated by a factor $\approx 3$
if compared with the mass calculated by a larger nuclear reaction
network (464 nuclei; \citealt{2015ApJ...808..164M}, and Paper I).
Indeed, taking into account this factor, the models
predicting a mass of $^{44}$Ti consistent with observations are
N16.3 and B18.3 (total mass $\approx 1.3\times 10^{-4}\,M_{\odot}$
and $\approx 1.4\times 10^{-4}\,M_{\odot}$, respectively).

\begin{table}
\caption{Parameters of the Gaussian functions best fitting the observed and
synthetic lines of $^{44}$Ti.}
\label{boggs_fit}
\begin{center}
\begin{tabular}{lcc}
\hline
\hline
        &   $v_{\rm rs}$   &   $\sigma$   \\
\hline
NuSTAR$^a$  &   $700^{+400}_{-400}$   &   $1043^{+565}_{-826}$  \\
N16.3       &   260   &    900  \\
B18.3       &   370   &   1100  \\
S19.8       &   840   &   1500  \\
\hline
\end{tabular}
\end{center}
\centerline{$^a$ NuSTAR observations (\citealt{2015Sci...348..670B}).}
\end{table}

NuSTAR observations around day $\approx 10000$ have revealed $^{44}$Ti
lines redshifted with Doppler velocities of $\approx 700\pm
400$~km~s$^{-1}$ (black line in the upper right panel of
Fig.~\ref{fig_Ti}; \citealt{2015Sci...348..670B}). From the models,
we synthesized the profile of $^{44}$Ti lines assuming a spectral
resolution of 900~eV in the range of wavelengths of interest (around
$[60-90]$~keV), and taking into account the Doppler shift and
broadening due to motion of titanium along the LoS. We fitted the
synthetic profiles with a Gaussian function, following the analysis
of NuSTAR observations (see upper right panel of Fig.~\ref{fig_Ti}).
Table~\ref{boggs_fit} reports the velocity of the peak of the
Gaussian, $v_{\rm rs}$, corresponding to the line redshift, and the
standard deviation, $\sigma$, corresponding to the Doppler broadening
after accounting for the NuSTAR spectral resolution at these energies.
In all the models, the synthetic $^{44}$Ti lines are redshifted.
However, model N16.3 produces a line profile (blue line in
the figure) with redshift much lower than observed ($\approx
260$~km~s$^{-1}$) and which is not compatible with the observations.
The other two models produce lines with redshift velocities of
$\approx 370$~km~s$^{-1}$ in B18.3 (red line in the figure) and
$\approx 840$~km~s$^{-1}$ in S19.8 (green line) which are both
consistent with the observations within the 90\% confidence area
(shaded gray region in the figure) for the position and width of
the Gaussian best-fitting the observations. In the three models,
the redshift originates from the initial asymmetry which assumes a
parameter $\alpha = v_{\rm up}/v_{\rm dw} = 1.5$ (see Table~\ref{sn_param});
this means that the lobe propagating away from the observer is
more energetic than that propagating toward the observer.

\subsection{Molecular structures in the evolved remnant}

The structure of innermost ejecta can be a direct probe of anisotropies
developed in the immediate aftermath of the SN. Recently, this
structure has been resolved by ALMA infrared observations of cold molecular
gas (\citealt{2017ApJ...842L..24A}) that have shown that the
distributions of carbon and silicon monoxide (CO and SiO) are
characterized by a torus-like structure perpendicular to the
equatorial ring.

\begin{figure*}[!t]
  \begin{center}
  \leavevmode
  \epsfig{file=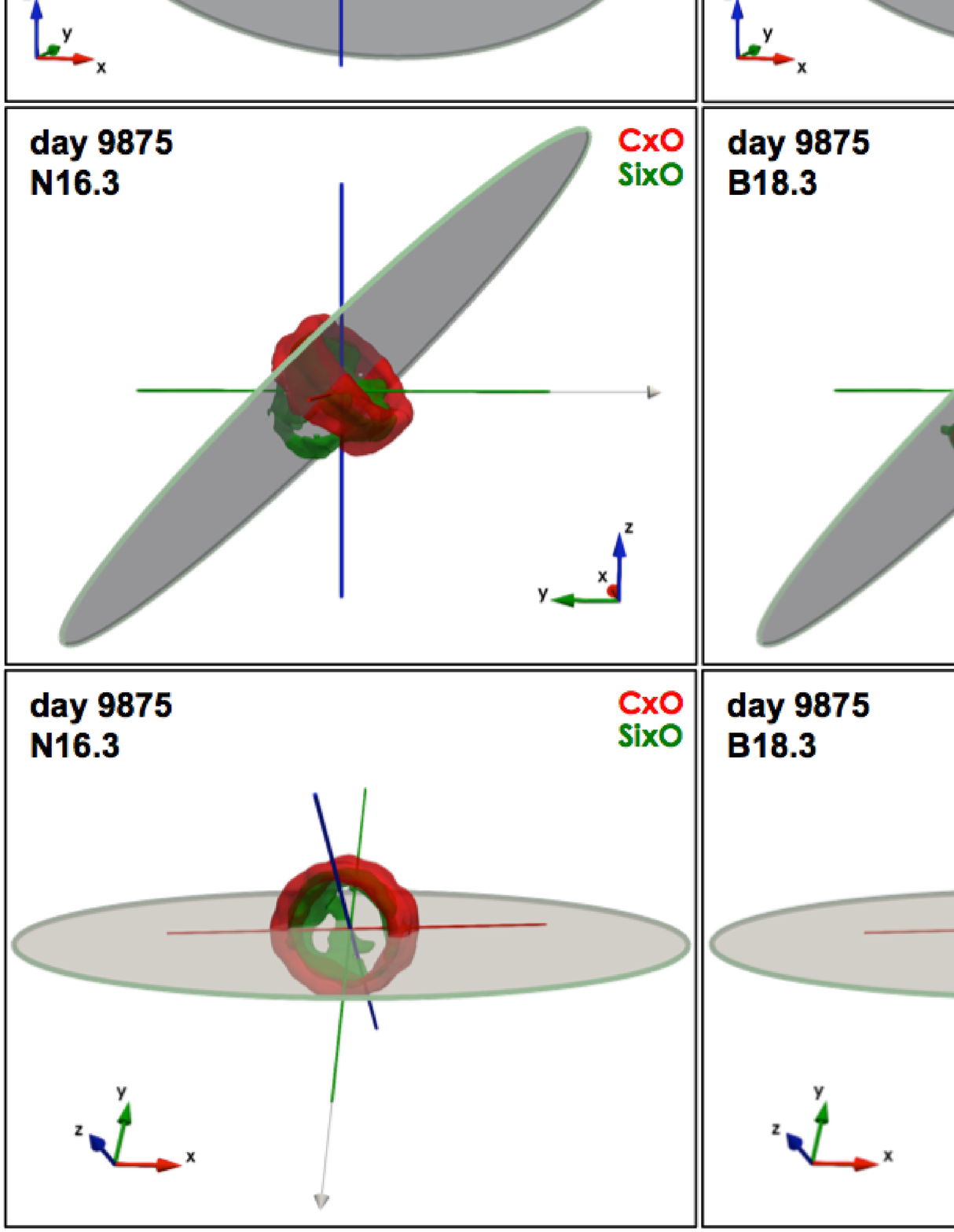, width=18.cm}
  \caption{Isosurfaces of the
  distributions of C$\times$O (red) and Si$\times$O (green) at day
  9875 from selected view angles for models N16.3 (left panels),
  B18.3 (center panels), and S19.8 (right panels). The semi-transparent
  isosurfaces correspond to a value of density which is at 50\% of
  the respective peak density (maximum density is $1.3\times
  10^{-19}$~g~cm$^{-3}$, $3.2\times 10^{-19}$~g~cm$^{-3}$, and
  $1.9\times 10^{-19}$~g~cm$^{-3}$ for Si$\times$O and $9.9\times
  10^{-20}$~g~cm$^{-3}$, $4.3\times 10^{-19}$~g~cm$^{-3}$, and
  $1.3\times 10^{-19}$~g~cm$^{-3}$ for C$\times$O, for models N16.3,
  B18.3, and S19.8 respectively). The gray arrow lying on the $y$
  axis indicate the position of the observer. The ring indicates
  the position of the dense inner ring. See on-line 
  Movie 6 for an animation of these data; a navigable 3D graphic
  of the C$\times$O and Si$\times$O distributions for the three
  models is available at the link {https://skfb.ly/6P7sI}.}
\label{fig_molecules}
\end{center}
\end{figure*}

Our models do not describe the formation of molecules. Nevertheless
we can evaluate the size scales and spatial distributions of CO and
SiO following the approach described in \cite{2017ApJ...842L..24A}.
More specifically, we used the modeled atomic density distributions
of $^{12}$C, $^{16}$O, and $^{28}$Si to derive the square root of
the products ($^{12}$C\,$\times $\,$^{16}$O) and ($^{28}$Si\,$\times
$\,$^{16}$O). These can be considered as proxies of CO and SiO.
In a forthcoming paper (Ono et al., in preparation), we will calculate
accurately the molecule formation in SN\,1987A using the MHD models
described here. Figure~\ref{fig_molecules} and the online Movie 6
show the distributions of $\sqrt{^{12}{\rm C}\times^{16}{\rm O}}$
(in the following C$\times$O) and $\sqrt{^{28}{\rm Si}\times^{16}{\rm
O}}$ (in the following Si$\times$O) at day 9875. Models N16.3 and
B18.3 show a dense toroidal structure for both C$\times$O and
Si$\times$O at 50\% of the peak density located around the center
of the explosion. The torus forms because the fast outflows
from the bipolar explosion pierce the outer shells of C, O and Si;
then this material is swept out by the outflows and progressively
accumulates at their sides, producing the torus-like feature. The
torus forms soon after the core-collapse; then it expands following
the evolution of the remnant. In model S19.8, this structure is
not clearly evident and incomplete shells of C$\times$O and Si$\times$O
develop around the center of the explosion.

The axis of the torus (when present) lies in the plane of the
equatorial ring in the direction of the initial large-scale asymmetry:
the clumps of early $^{56}$Ni, $^{56}$Fe, and $^{44}$Ti, all travel
along this axis. Therefore, the models show that the initial explosion
asymmetry required to fit the line profiles of [Fe\,II] 26 $\mu$m
and 18 $\mu$m few years after the SN predicts a toroidal structure
of C$\times$O and Si$\times$O, in models N16.3 and B18.3, which is
oriented in the same way as the dense structure of CO and SiO
inferred from observations (\citealt{2017ApJ...842L..24A}). We note
that the dense structures of CO and SiO observed with ALMA are quite
irregular and their torus-like shape (as suggested by observations)
relies just on few plumes of dense CO and SiO. Nevertheless, the
observations show that these plumes lie on a preferential plane
whose orientation coincides with that of the torus-like feature of
models N16.3 and B18.3.

\begin{figure*}[!t]
  \begin{center}
  \leavevmode
  \epsfig{file=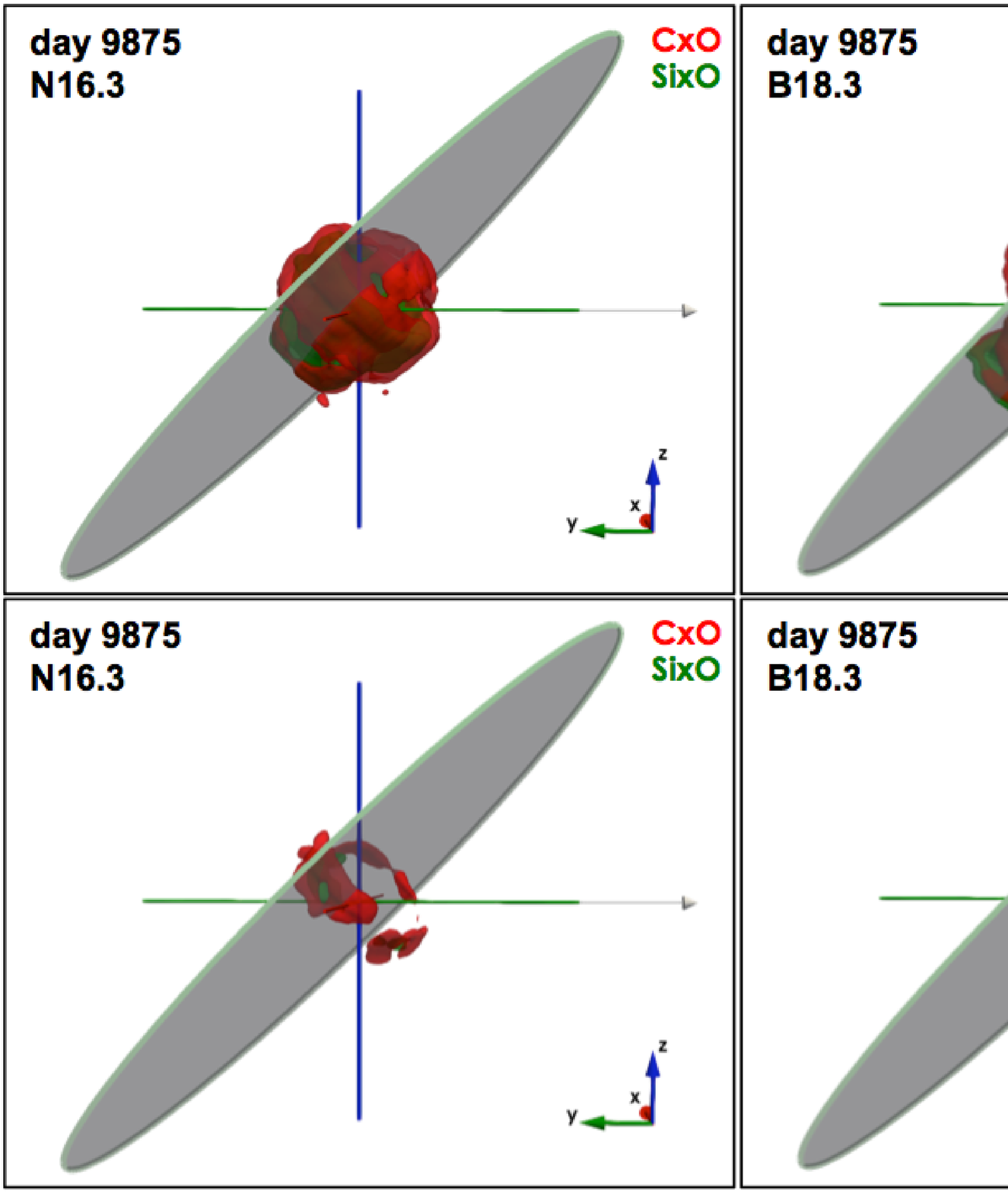, width=18.cm}
  \caption{As in Fig.~\ref{fig_molecules} but for semi-transparent
  isosurfaces corresponding to a value of density which is at 30\%
  (upper panels) and 70\% (lower panels) of the respective peak.}
\label{fig_molecules_th}
\end{center}
\end{figure*}

\begin{figure*}[!t]
  \begin{center}
  \leavevmode
  \epsfig{file=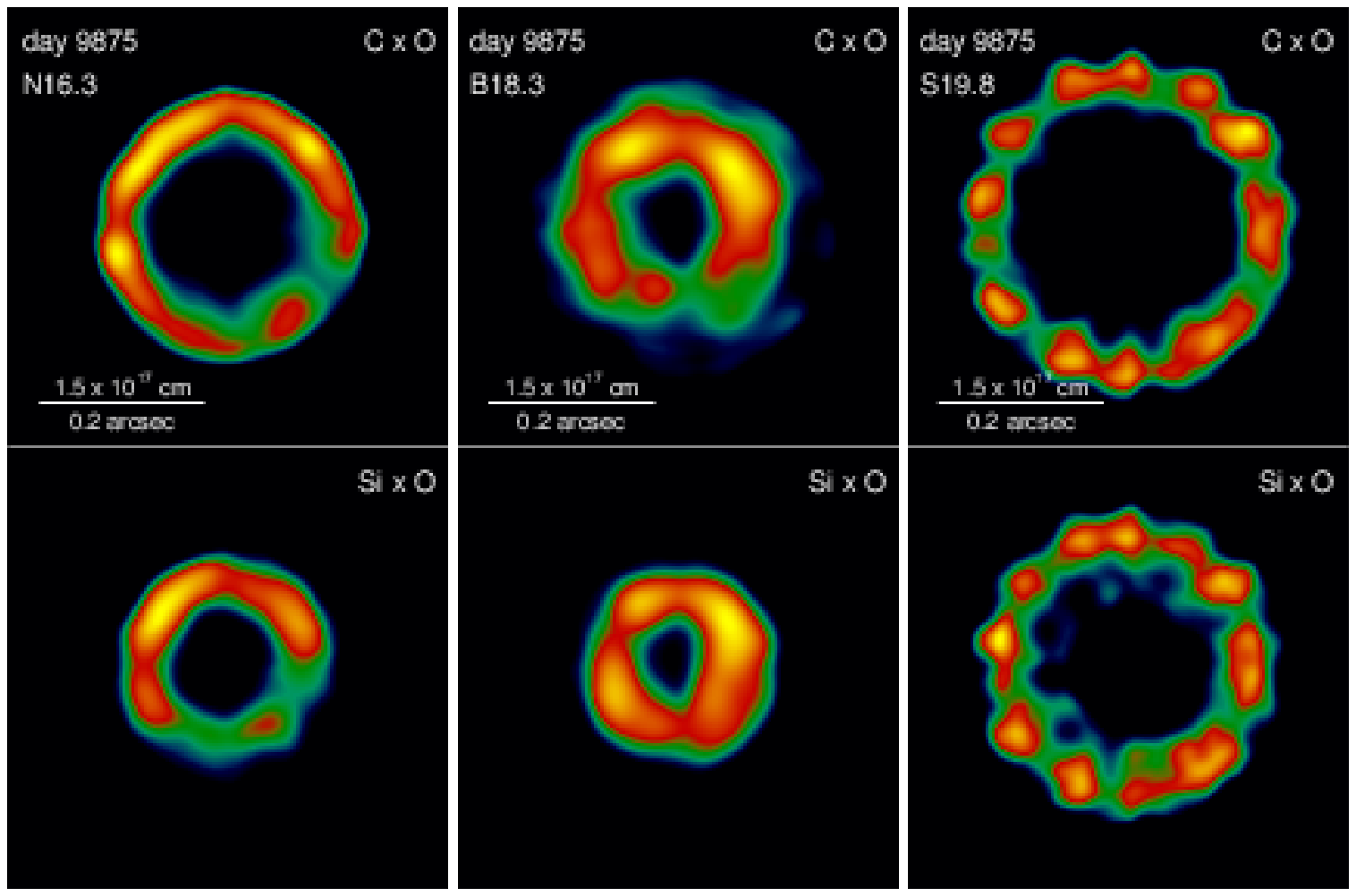, width=18.cm}
  \caption{Maps of C$\times$O (upper panels) and Si$\times$O (lower panels)
  integrated along the axis of the torus-like structure at day 9875
  for models N16.3, B18.3, and S19.8. The yardstick in the upper
  panels indicates the length scale in cm and in arcsec, assuming
  a distance of 51.4~kpc.}
\label{fig_molecules_torus}
\end{center}
\end{figure*}

Fig.~\ref{fig_molecules_th} shows the distributions of C$\times$O
and Si$\times$O for isosurfaces corresponding to densities at 30\%
and 70\% of the peak. At 30\% of the peak density, the central
deficit of density is fully enclosed into a shell of C$\times$O and
Si$\times$O in all models. At 70\% of the peak, the torus-like
structure evident at 50\% of the peak in models N16.3 and B18.3 is
still clearly visible, whilst an irregular shell characterizes the
distribution of C$\times$O and Si$\times$O in model S19.8.

A more quantitative comparison between the modeled and the observed
structures can be done by comparing the typical size scales
inferred from observations and obtained with the three models around
day $\approx 10000$. Figure~\ref{fig_molecules_torus} shows maps
of C$\times$O and Si$\times$O derived by integrating their density
distributions along the axis of the toroidal structure at day 9875
for the three models. The modeled distributions are more regular
than those observed; in particular both C$\times$O and Si$\times$O
exhibit clear and regular ring-like distributions at odds with the
observations that show more clumpy and irregular ring-like morphologies
for CO and SiO distributions. This is most likely due to the idealized
initial asymmetric explosion prescribed in our simulations.
Nevertheless, the large-scale morphology of the modeled structures
and their typical length scales are very similar to those observed.

\begin{figure*}[!t]
  \begin{center}
  \leavevmode
  \epsfig{file=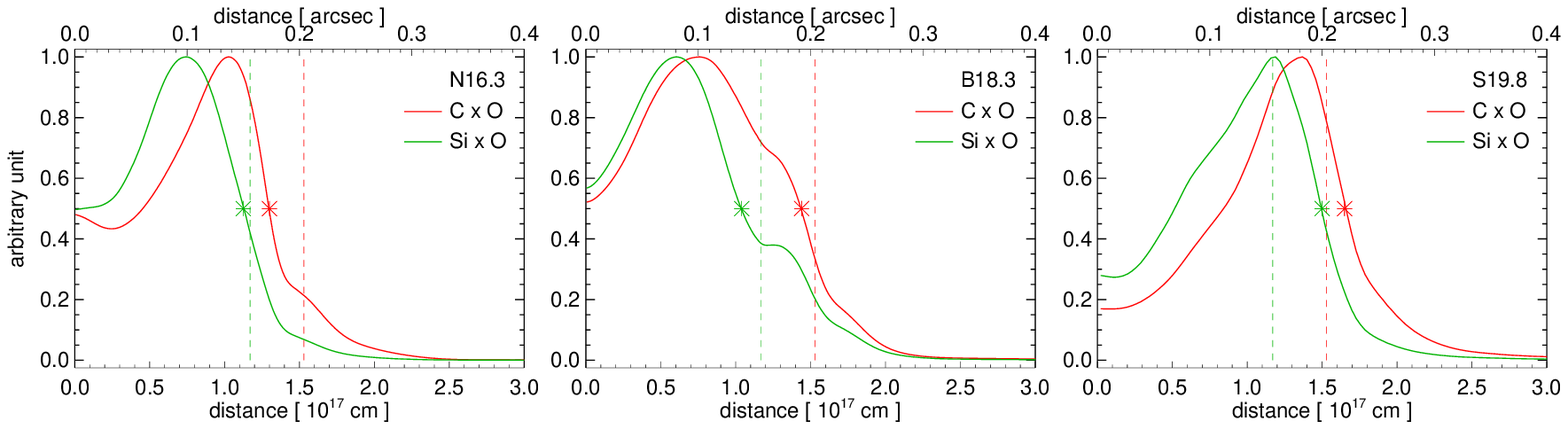, width=18.cm}
  \caption{Average radial distribution of C$\times$O (red curve) and
  Si$\times$O (green) from the axis of the torus-like structure at day
  9875 for models N16.3, B18.3, and S19.8. Symbols indicate the
  extension of the C$\times$O (red stars) and Si$\times$O (green
  stars) structures at 50\% of the peak of the profiles; vertical
  dashed lines show the maximum extension of CO (red) and SiO (green)
  inferred from ALMA observations (\citealt{2017ApJ...842L..24A}).}
\label{fig_molecules_quant}
\end{center}
\end{figure*}

From Fig.~\ref{fig_molecules_torus} we note that, in all models,
C$\times$O presents a maximum extension larger than Si$\times$O as
also found in the observed distributions of CO and SiO
(\citealt{2017ApJ...842L..24A}). This is more clear by inspecting
Fig.~\ref{fig_molecules_quant} which shows the angle-averaged radial
distributions of C$\times$O and Si$\times$O derived from the maps
in Fig.~\ref{fig_molecules_torus}. In all the models, C$\times$O
and Si$\times$O are spatially distinct as in the observations.
Models N16.3 and B18.3 predict a size of C$\times$O and Si$\times$O
(see Table~\ref{tab_molec}) of the same order of magnitude of the
observed maximum extension of CO and SiO ($\approx \pm 1.53\times
10^{17}$~cm and $\approx \pm 1.17\times 10^{17}$~cm, respectively;
\citealt{2017ApJ...842L..24A}). Model S19.8 predicts a size of C$\times$O
and Si$\times$O significantly larger than observed.

\begin{table}
\caption{Maximum extension of CO and SiO inferred from the observations
and derived from the models.}
\label{tab_molec}
\begin{center}
\begin{tabular}{lcc}
\hline
\hline
        & $R_{\rm C\times O}$ & $R_{\rm Si\times O}$ \\
        & [$10^{17}$~cm]      & [$10^{17}$~cm] \\
\hline
ALMA$^a$ &  1.53  &  1.17 \\
N16.3    &  1.30  &  1.12  \\
B18.3    &  1.44  &  1.04 \\
S19.8    &  1.65  &  1.50  \\
\hline
\end{tabular}
\end{center}
\centerline{$^a$ ALMA observations (\citealt{2017ApJ...842L..24A}).}
\end{table}

Despite the idealization of the initial asymmetry of the SN explosion,
we found that, in models N16.3 and B18.3, the asymmetry and its
orientation, which are necessary to reproduce the redshift and
broadening of observed [Fe\,II] lines during the first two years after the SN
(\citealt{1990ApJ...360..257H}) and the redshift of $^{44}$Ti observed
at day $\approx 10000$, naturally produce a toroidal dense structure
in the distributions of C$\times$O and Si$\times$O (proxies of CO
and SiO molecules) at day $\approx 10000$ with position, orientation,
and size similar to those of observed CO and SiO distributions
(\citealt{2017ApJ...842L..24A}). We conclude that all ejecta anisotropies
discussed above originate from the same large-scale asymmetry in
the SN explosion.

\subsection{Neutron star kick}
\label{starkick}

The highly asymmetric explosion is expected to cause momentum
transfer to a newly formed compact object possibly produced by
SN\,1987A (e.g. \citealt{2013A&A...552A.126W}): the proto-neutron
star and the center of gravity of the ejecta are expected to move
in opposite directions due to momentum conservation. In the case
of the initial asymmetry required by our simulations to reproduce
the observations of the SN and the SNR, the kick is mainly
regulated by the parameter $\alpha$ of the asymmetry which also
determines the shift of the centroid of [Fe\,II] lines about 1-2
years after the SN event. The parameter is well constrained by the
comparison between the modeled and the observed shifts (see
Sect.~\ref{sec_sn_evol}). Thus, the kick was estimated by assuming
momentum conservation as in \cite{2013A&A...552A.126W} (see also
Paper I): it is calculated by imposing the initial total momentum
equal zero. We found that all the three cases investigated predict
that the compact object would have received a kick velocity of
$\approx 300$~km~s$^{-1}$ toward the observer in the north (see the
lightblue arrow in Fig.~\ref{fig_orient}; see also Paper I).

We note that the initial asymmetry may have been more complex
than that modeled here, possibly producing plumes and fingers in
different directions. In fact, as discussed in Sect.~\ref{sec_sn_evol},
our models do not describe, for instance, the isolated fast clump
of iron responsible for the spectral feature redshifted at $v\approx
4000$~km~s$^{-1}$ in [Fe\,II] lines (see upper right panel of
Fig.~\ref{fig1}). That feature may be a signature of a structure
of the initial asymmetry more complex than modeled here. One may
ask, for instance, if the explosion was not strictly bipolar but
had dominant outflows separated by less than 180$^{\rm o}$. In our
models, we considered simple idealized forms for the asymmetry,
with a bipolar structure. The fact that one of our models is able
to broadly reproduce the observed line profiles suggests that two
dominant outflows were produced by the explosion, a prominent one
redshifted and the other blueshifted, separated by an angle which
was not much lower than 180$^{\rm o}$.  Other smaller scale structures
may have been produced at the collapse, but our simulations show
that the bulk of the explosion kinematics was dominated by the
bipolar outflows. Nevertheless, we note that the predicted kick
velocity might be a lower limit to the actual velocity. In fact,
the high velocity spectral feature in the red wing of [Fe\,II] lines
suggests that the bulk of ejecta (not only iron) might present more
mass traveling away from the observer. In this case, our models
might underestimate the momentum of ejecta traveling away from the
observer and, consequently, the momentum of the compact object
traveling toward the observer.

Our estimate for the neutron star kick is in good agreement with
previous studies (\citealt{2000ApJS..127..141N, 2013A&A...552A.126W,
2017IAUS..331..148J}). Early studies pointed out that a highly
asymmetric explosion required for SN\,1987A would have produced a
kick velocity of the same order of magnitude as found here
(\citealt{2000ApJS..127..141N}), although the authors suggested that the
proto-neutron star should move toward the observer in the south
because they have assumed that the initial asymmetry was aligned
with the axis of the equatorial ring. More recently, the comparison
of the iron distribution derived in a 3D neutrino-driven explosion
model for a $15 M_{\odot}$ star appropriate for SN\,1987A
(\citealt{2017IAUS..331..148J}) with 3D maps of silicon and iron
derived from spectral and imaging observations of SN\,1987A with
HST/STIS and VLT/SINFONI (\citealt{2016ApJ...833..147L}) suggested
a kick velocity of the same order of magnitude toward the observer
in the north, but with an angle with respect to the LoS slightly
smaller than in our models (compare Fig.~4 in \citealt{2017IAUS..331..148J}
with Fig.~\ref{fig_orient} in this paper).

Finally, our model predictions are also in good agreement with
recent findings from the analysis of ALMA observations of dust in
the SN\,1987A ejecta (\citealt{2019ApJ...886...51C}). The analysis
has shown a dust blob in ALMA images which is slightly offset along
the north-east direction from the estimated position of Sk $-69^{\rm
o}\, 202$. \cite{2019ApJ...886...51C} have interpreted the blob as
dust heated by the compact object and estimated its kick velocity
to be $\approx 700$~km~s$^{-1}$. The estimate of the kick was done
by considering the age of SN\,1987A and the offset between the
brightest part of the blob (interpreted as an early pulsar wind
nebula) and the position of Sk $-69^{\rm o}\, 202$. However,
in the Crab Nebula, the position of the compact object and that of
the brightest part of the pulsar wind nebula are not coincident
(\citealt{2000ApJ...536L..81W}). Thus some uncertainty in the
determination of the offset and of the kick velocity should be
considered. Nevertheless, the predicted position of the neutron
star and the lower limit on its kick velocity derived with our
favorite model B18.3 are both fully consistent with the interpretation
proposed by \cite{2019ApJ...886...51C}.

\section{Summary and conclusions}
\label{conclusion}

We investigated the evolution of the remnant of SN\,1987A with the
aim of unveiling the link between the radiative and dynamical
properties of the remnant and the properties of the parent SN
explosion and structure of the progenitor star. To this end, we
developed a comprehensive 3D hydrodynamic model which describes the
long-term evolution of SN\,1987A from the onset of the SN to the
development of its remnant, accounting for the pre-SN structure of
the progenitor star. The model was realized by coupling a 3D model
of a core-collapse SN (\citealt{2013ApJ...773..161O} and Paper I)
with a 3D model of a SNR (\citealt{2015ApJ...810..168O,
2019A&A...622A..73O}).

The SN model has the relevant physics required to describe the
evolution in the immediate aftermath of the core-collapse (see
Sect.~\ref{SNmodel}): the explosive nucleosynthesis through a nuclear
reaction network, the feedback of nuclear energy generation, the
energy deposition due to radioactive decays of isotopes synthesized
in the SN, the effects of gravity (both self-gravity and gravitational
effects of the central proto-neutron star), and the fallback of
material on the proto-neutron star. Aspherical explosions (as
suggested by observations; \citealt{2016ApJ...833..147L}) were
simulated to mimic the global anisotropies developing in jet-driven
SNe, or by convection in the neutrino heating layers and the SASI
(Paper I).

The 3D simulations of the SN were initialized from pre-SN stellar
models available in literature (see Sect.~\ref{stellar_prog}): a
$16.3 M_\odot$ BSG (model N16.3; \citealt{1990ApJ...360..242S}), a
$18.3 M_\odot$ BSG (B18.3; \citealt{2018MNRAS.473L.101U}), and a
$15.9 M_\odot$ RSG (S19.8; \citealt{2016ApJ...821...38S}). Model
B18.3 results from the merging of two massive stars and reproduces
most of the observational constraints of the progenitor star of
SN\,1987A: the red-to-blue evolution, its lifetime, the total mass
and the position in the Hertzsprung-Russell diagram at collapse
(\citealt{2018MNRAS.473L.101U}).

We then used the output from the SN simulations (about 20 hours
after the core-collapse) to start 3D MHD simulations, describing
the subsequent expansion of the ejecta through the inhomogeneous
pre-SN environment. The SNR model includes the relevant physics to
describe the evolution of the remnant and its interaction with the
inhomogeneous ambient environment (see Sect.~\ref{SNRmodel}): a
plausible configuration of the ambient magnetic field, the deviations
from temperature-equilibration between electrons and ions, and the
deviations from equilibrium of ionization of the most abundant ions.
The inhomogeneous pre-SN ambient environment was described as a
nebula consisting of a dense ring surrounded by a H\,II region with
3D geometry compatible with observations (e.g.
\citealt{2005ApJS..159...60S, 2013MNRAS.429.1324S}). The simulations
cover 50 years, thus describing the evolution of SN\,1987A since
the SN and providing some hints on its evolution until 2037.
Current and future observations will help in unveiling the structure
and geometry of the medium beyond the dense equatorial ring (e.g.
\citealt{2019arXiv191009582L}) and will provide important constraints
for the models.

We explored different explosion energies and asymmetries, and
different orientations of the asymmetry with respect to the LoS,
to identify the physical and geometrical properties of the SN
explosion which led to the shifts and broadening of [Fe\,II] lines
in early observations (e.g. \citealt{1990ApJ...360..257H,
1994ApJ...427..874C, 1995A&A...295..129U}). Furthermore, we explored
several density structures of the nebula around the parameters found
in previous works (\citealt{2015ApJ...810..168O, 2019A&A...622A..73O})
and within the range of values inferred from observations (e.g.
\citealt{2012ApJ...752..103D, 2005ApJS..159...60S}), searching for
parameters which best reproduce X-ray lightcurves and morphology
of the remnant observed in the last 30 years (e.g.
\citealt{2006A&A...460..811H, 2012A&A...548L...3M, 2013ApJ...764...11H,
2015ApJ...810..168O, 2016ApJ...829...40F}), and the observed late
distribution of metal rich ejecta (e.g. \citealt{2015Sci...348..670B,
2017ApJ...842L..24A}).

The best match with observations was found for models with a highly
asymmetric SN explosion with a total energy at the shock breakout
of $\approx 2\times 10^{51}$~erg (see Table~\ref{sn_param}). The
energy is slightly higher than that estimated in a previous work
(\citealt{2015ApJ...810..168O}), but still within the range of
values expected for SN\,1987A. For all the progenitor stars
investigated, the observations require that a significant fraction
of energy was channeled along an axis almost lying in the plane of
the central ring, roughly in the direction of the LoS but with an
offset of $40^{\rm o}$ (see Fig.~\ref{fig_orient}); the projection
of this axis in the plane of the sky is offset by $15^{\rm o}$ from
the northsouth axis. Such large asymmetries may result from a
combination of low-mode convection in the neutrino heating layers
and strong axial sloshing of the shock due to the SASI (e.g.
\citealt{2003ApJ...584..971B}); alternatively, extreme asymmetries
may develop in magnetic, jet-driven SNe (e.g.
\citealt{1999ApJ...524L.107K}) or in single-lobe SNe (e.g.
\citealt{2005ApJ...635..487H}).

\begin{table}
\caption{Comparison of model results and observations; the Table
summarizes the observational constraints (for the progenitor star,
SN, and SNR) satisfied (\cmark) or not satisfied (\xmark) by the
three best-fit models.}
\label{models_observ}
\begin{center}
\begin{tabular}{llccc}
\hline
\hline
                 &                         &  N16.3   &   B18.3   &   S19.8\\
\hline
progenitor       &  red-to-blue evolution     &  \xmark  &   \cmark  &   \xmark  \\
star             &  lifetime of BSG phase     &  \xmark  &   \cmark  &   \xmark  \\
                 &  mass                      &  \cmark  &   \cmark  &   \cmark  \\
                 &  radius                    &  \cmark  &   \cmark  &   \xmark  \\
                 &  $\log T_{\rm eff}$        &  \cmark  &   \cmark  &   \xmark  \\
                 &  $\log L/L_{\odot}$        &  \cmark  &   \cmark  &   \cmark  \\
\hline
SN               &  explosion energy          &  \cmark  &   \cmark  &   \cmark  \\
                 &  early $^{56}$Ni clumps    &  \cmark  &   \cmark  &   \cmark  \\
                 &  [Fe\,II] line profiles    &  \xmark  &   \cmark  &   \xmark  \\
\hline
SNR              &  X-ray lightcurves         &  \cmark  &   \cmark  &   \xmark  \\
                 &  X-ray morphology   \\
                 &  ~~~~and size              &  \xmark  &   \cmark  &   \xmark \\
                 &  $^{44}$Ti line profiles   &  \xmark  &   \cmark  &   \cmark  \\
                 &  CO and SiO morphology \\
                 &  ~~~~and size              &  \cmark  &   \cmark  &   \xmark \\
\hline
\end{tabular}
\end{center}
\end{table}

We found that the structure of the progenitor star and the initial
asymmetry of the SN both play a central role in determining the
mixing of $^{56}$Ni to speeds exceeding 3000~km~s$^{-1}$, and the
observed high redshifts of [Fe\,II] lines around day 400 and $^{44}$Ti
lines around day 10000. Among the models investigated, we found
that B18.3 is the only one satisfying the observational constraints
for the progenitor star, the SN, and the SNR (see
Table~\ref{models_observ}).  In fact, this model is the only one
able to reproduce the mixing of ejecta which leads to the observed
velocity distributions of $^{56}$Ni and $^{56}$Fe. This is possible
because the C+O and He layers in this model are significantly less
extended and less dense than in the others. As a result, nickel
(and heavy elements formed in the innermost layers of the remnant)
can penetrate efficiently through the helium and hydrogen layers
producing the observed mixing and velocities.

The imprint of the progenitor star is also evident in the X-ray
observations of the SNR. The X-ray lightcurves of SN\,1987A can be
reproduced by the two BSG (N16.3 and B18.3) but not by the RSG
progenitor (S19.8). This is due to the structure of the extended
hydrogen envelope of the RSG which causes a significant deceleration
of the blast wave before the shock breakout, so that, when the blast
travels through the CSM, it has a velocity much smaller than in the
other two models examined. Furthermore, the differences in the
structure of the progenitor stars make B18.3 the model best reproducing
also the size and morphology of the X-ray emitting remnant during
its evolution, whilst the other two models fail in reproducing at
the same time fluxes and size of the X-ray source. We conclude that
the dynamical and radiative properties of the remnant require a
progenitor BSG that resulted from the merging of two massive stars
before the collapse.

The observed distribution of cold molecular gas (e.g. CO and SiO)
is also consistent with the same progenitor star and explosion
asymmetry needed to reproduce the high redshifts and broadenings
of [Fe\,II] and $^{44}$Ti lines, and the X-ray lightcurves and
morphology of the remnant. This suggests that all these features
are signatures of the same process associated with the SN.

As a result of the highly asymmetric explosion, we estimated that
a compact object possibly produced by SN\,1987A would have received
a kick velocity of $\approx 300$~km~s$^{-1}$ toward the observer
in the north (lightblue arrow in Fig.~\ref{fig_orient}; see
Sect.~\ref{starkick}).  Currently the newly formed neutron star (if
any) has not been detected yet. This is probably due to heavy
absorption of its emission caused by a high local absorbing column
density along the LoS (\citealt{2015ApJ...810..168O, 2018ApJ...864..174A}).
Future detection of the neutron star in SN\,1987A will provide a
strong constraint on the initial large-scale asymmetry and on the
mechanism of neutron star kick.

Our long-term simulations provided a way to link the remnant of
SN 1987A to its parent SN explosion and to the internal structure
of its progenitor star. Extending the present work to other young
SNRs will permit advances in resolving pending questions about the
physical processes associated with SNe and the final stages of stellar
evolution.

\begin{acknowledgements}
We thank the anonymous referee for useful suggestions that have
allowed us to improve the paper. We acknowledge that the results
of this research have been achieved using the PRACE Research
Infrastructure resource Marconi based in Italy at CINECA (PRACE
Award N.2016153460). Additional numerical computations were carried
out complementarily on XC40 (YITP, Kyoto University), Cray XC50
(Center for Computational Astrophysics, National Astronomical
Observatory of Japan), HOKUSAI (RIKEN). This research also used
computational resources of the K computer provided by the RIKEN
Center for Computational Science through the HPCI System Research
project (Project ID:hp180281). The FLASH code, used in this work,
is developed by the DOE-supported ASC/Alliance Center for Astrophysical
Thermonuclear Flashes at the University of Chicago (USA). The PLUTO
code is developed at the Turin Astronomical Observatory (Italy) in
collaboration with the Department of General Physics of Turin
University (Italy) and the SCAI Department of CINECA (Italy). SO
would like to thank H.-T. Janka for helpfull discussions on this
work and A. Mignone for his support with the PLUTO code. SO, MM,
GP, FB acknowledge financial contribution from the agreement ASI-INAF
n.2017-14-H.O, and partial financial support by the PRIN INAF 2016
grant ``Probing particle acceleration and $\gamma$-ray propagation
with CTA and its precursors''. This work is supported by JSPS
Grants-in-Aid for Scientific Research ``KAKENHI'' Grant Numbers
JP26800141 and JP19H00693.  SN and MO wish to acknowledge the support
from the Program of Interdisciplinary Theoretical \& Mathematical
Sciences (iTHEMS) at RIKEN (Japan). SN also acknowledges the support
from Pioneering Program of RIKEN for Evolution of Matter in the
Universe (r-EMU). The navigable 3D graphics have been developed
in the framework of the project 3DMAP-VR (3-Dimensional Modeling
of Astrophysical Phenomena in Virtual Reality;
\citealt{2019RNAAS...3..176O}) at INAF-Osservatorio astronomico di
Palermo
(http://cerere.astropa.unipa.it/progetti\_ricerca/HPC/3dmap\_vr.htm) and
uploaded on the Sketchfab platform.

\end{acknowledgements}

\bibliographystyle{aa}
\bibliography{references}

\end{document}